\documentclass[twocolumn,prb,superscriptaddress]{revtex4-2} 

\usepackage{graphicx}

\usepackage[utf8]{inputenc}
\usepackage{sidecap}
\usepackage{caption, subcaption}
\usepackage{units}
\DeclareUnicodeCharacter{2212}{-}
\usepackage[T1]{fontenc}
\usepackage{textgreek}
\usepackage{color,soul} 
\usepackage{comment}
\usepackage{float} 
\usepackage{lipsum}
\usepackage{physics}
\usepackage{array}
\usepackage{url}
\usepackage{amsmath}
\usepackage{upgreek}


\date{}
\newcommand{\cmt}{Cd$_{0.7}$Mg$_{0.3}\!$Te }
\newcommand{\kp}{{\bf k$\cdot$p}\ }
\newcommand{\Pp}{{\bf P$\cdot$p}\ }
\newcommand{\cdteQW}{CdTe/(Cd,Mg)Te}

\begin{document}
\preprint{APS/123-QED}

\title{Experiment and \kp  analysis of  the  luminescence from modulation-doped CdTe/(Cd,Mg)Te quantum
wells at magnetic field}
\author{W. Solarska}
\affiliation{Faculty of Physics, University of Warsaw, L. Pasteura 5, 02-093 Warsaw, Poland}

\author{M. Grymuza}
\affiliation{Faculty of Physics, University of Warsaw, L. Pasteura 5, 02-093 Warsaw, Poland}

\author{M. Kubisa}
\affiliation{Wrocław University of Science and Technology, Department of Experimental Physics,
Wybrzeże Wyspiańskiego 27, 50-370 Wrocław, Poland}

\author{K. Ryczko}
\affiliation{Wrocław University of Science and Technology, Department of Experimental Physics,
Wybrzeże Wyspiańskiego 27, 50-370 Wrocław, Poland}

\author{P. Pfeffer}
\affiliation{Institute of Physics, Polish Academy of Sciences, Aleja Lotników 32/46, PL-02-668 Warszawa, Poland}

\author{K. P.  Korona}
\affiliation{Faculty of Physics, University of Warsaw, L. Pasteura 5, 02-093 Warsaw, Poland}
 
\author{K.~Karpierz}
\affiliation{Faculty of Physics, University of Warsaw, L. Pasteura 5, 02-093 Warsaw, Poland}

\author{D.~Yavorskiy}
\affiliation{Institute of High Pressure Physics, Polish Academy of Sciences, ul. Sokołowska 29/37, 01-142 Warsaw, Poland}
\affiliation{Institute of Physics, Polish Academy of Sciences, Aleja Lotników 32/46, PL-02-668 Warszawa, Poland}
\affiliation{CENTERA, CEZAMAT, Warsaw University of Technology, ul. Poleczki 19, 02-822 Warsaw, Poland}

\author{Z.~Adamus}
\affiliation{Institute of Physics, Polish Academy of Sciences, Aleja Lotników 32/46, PL-02-668 Warszawa, Poland}
\affiliation{International Research Centre MagTop, Institute of Physics, Polish Academy of Sciences, al. Lotników 32/46, 02-668 Warsaw, Poland}
 
\author{T.~Wojtowicz}
\affiliation{International Research Centre MagTop, Institute of Physics, Polish Academy of Sciences, al. Lotników 32/46, 02-668 Warsaw, Poland}


\author{J. Łusakowski}
\email[]{jerzy.lusakowski@fuw.edu.pl}
\affiliation{Faculty of Physics, University of Warsaw, L. Pasteura 5, 02-093 Warsaw, Poland}

\date{\today}

\begin{abstract}
In spite of a large quantity  of papers devoted to the mangetoluminescence from \cdteQW\, quantum wells there have been no attempts to analyze it on the basis of the band-structure calculations. This has been proposed in the  present paper.  Samples containing one or ten CdTe quantum wells with \cmt barriers are grown by a molecular beam epitaxy on a semi-insulating GaAs substrate. Each well is modulation-doped
with iodine donors which leads to the creation of a two-dimensional electron gas  in the wells.  Polarization-resolved ($\sigma^+/\sigma^-$) photoluminescence spectra are measured at liquid helium temperatures  and magnetic fields up to 9~T. 
The results are interpreted on the basis of  calculations of the energy of  Landau levels in the  conduction  and valence bands. In the latter case,  we use the Luttinger Hamiltonian  while the conduction band is described within a three-level $\mathbf{k} \cdot\mathbf{p}$ model. Both models, originally formulated for bulk materials,  are adapted for two-dimensional structures.  We have found that the majority of all observed  transitions is well reproducede by this theory. However, some strong transitions are not which allows us to propose an enlarged scheme of selection rules of the photoluminescence transitions resulting from mixing of the conduction and valence bands.  We observe transitions involving Landau levels in the valence band with the index up to 7. To understand the origin of occupation  with photoexcited holes of these levels, lying  deep in the valence band, we carry  out time-resolved measurements which show that the photoexcited barrier is a source of long-lived holes tunneling into the quantum wells.  Calculations of the conduction band electron effective $g$-factor  show its strong variation with   the electron's energy and the external magnetic field. 
\end{abstract}

\maketitle

\section{Introduction}{\label {Introduction}}

The  technology of growth of \cdteQW\, quantum wells (QWs)~\cite{AWaag_1993, AWaag_1994, DGerthsen_1994, TWojtowicz_1997}  has been currently developed  to a high level and fabricated structures enable one to observe fundamental phenomena like the integer and fractional quantum Hall effects~\cite{BAPiot_2010} or fluctuations of the quantum conductance~\cite{MCzapkiewicz_2012}.   Studies of the photoluminescence (PL) from the \cdteQW\,  QWs at high magnetic field ($B$)   and low temperatures are an important part of a broad research on optical-properties of these structures. They were initiated soon after the first fabrication of this kind of QWs within a broader contex of research on different non-magnetic QWs based on II-VI materials~\cite{BKuhn_1993,AWaag_1994,DGerthsen_1994,TWojtowicz_1997}.

 Optical properties   of QWs depend on the  carrier concentration in a two-dimensional electron gas (2DEG) and, in particular, the PL spectrum evolves with its increase. We skip here a discussion of effects related to a two-dimensional hole gas because in  the present work we deal only with photoexcited holes of a low concentration. In empty QWs, one observes excitons, in lightly-doped samples charged excitons appear~\cite{KKheng_1993,PKossacki_2003} and in highly-doped samples the spectrum shows the Fermi-edge singularity (FES)~\cite{PHawrylak_1991,GColi_1997,VHuard_2000,RASuris_2001,YImanaka_2005, DAndronikov_2005}. An interaction of excitons and triones mediated by a 2DEG was considered in~\cite{RASuris_2001, CRLPNJeukens_2002,JTribollet_2003}. A three-particle excitation, different from a trion, was reported and analyzed in a few papers~\cite{DRYakovlev_1997,VPKochereshko_1997,VPKochereshko_1998}; in this case an incident photon creates an exciton and simultaneously excites an electron from one Landau level (LL) to another. 
The PL generated by the mentioned above  excitations is accompanied by PL due to other transitions, like different kinds of transitions between free electrons and holes bound on acceptors and acceptor-like centres, bound electrons to bound holes, biexcitons or phonon-assisted recombination. This picture is further enriched  by the dependence of   the PL spectrum on  the way of excitation (with a below- or over-barrier photon energy), an influence of lowering of the structure's  symmetry at the interfaces and external fields. 

A vividly explored  direction of research on \cdteQW\, QWs was related to studies of the dynamics of excitons and trions with a particular attention paid to spin coherence~\cite{RBratschitsch_2006,ZChen_2007,EAZhukov_2007,JHVersluis_2009,CPhelps_2009,GMoody_2014,MSalewski_2017,SVPoltavtsev_2017, ANKosarev_2019,SVPoltavtsev_2020,SVPoltavtsev_2020A}.

In the present paper, we concentrate on the evolution of the continuous wave  PL spectrum of \cdteQW\, QWs with $B$  applied perpendicular to the QWs plane and its theoretical analysis with a model based on \kp calculations.  The  description of both the conduction band (CB)  and the valence band (VB) in the QWs  starts with the Hamiltonians of bulk materials with additional self-consistent potential created by doping and the 2DEG and with taking into account boundary conditions at the interfaces. The problem thus formulated is analitically untreatable and only numerical solutions can be obtained.  In the case of the CB, we apply a 3-level  \kp model which changes into a 3-level \Pp model at a non-zero $B$. Then,  in particular,  a mixing of harmonic  oscillator wave functions occurs  within a  LL of a given index.  Also, mixing and non-parabolicity of bands leads to $B$-dependence of the electron's effective $g$-factor,  $g^{*}$.

The VB is described by a model  which is based on the Luttinger Hamiltonian for the four-fold degenerate $\Gamma_8$ band,  applied to the   QWs.  Solutions of appropriate equations show a complicated structure of Landau levels in the VB of a zinc-blende QW in the magnetic field. Here, as it is in the case of a non-parabolic CB, a LL with a given  index $n_v$ contains harmonic oscillator functions of different indexes, in a way more complicated than in the case of the CB, as it will be shown below.

Mixing of different oscillator functions within one LL in both bands   influence  selection rules for polarization-resolved transitions leading to an essentially reacher scheme that could be expected in a simplified model based on a  parabolic band approximation. Nevertheles, it has to be underlined that the presented analysis of transitions, although quite satisfactory at the outcome, can be further developed to include more detailed description of the band structure.

Interpretation of experimental data of  the magneto-PL  from QWs with theoretical models based on the band calculations is by no means a new idea, but it seems   not to be applied in the past to   \cdteQW\, QWs and thus our paper is a new contribution to this field. The number of papers in which the band structure of semiconducting QWs was calculated is huge and it is neither possible to cover this subject in this introduction nor to give a very general description of the results  because of pecularities of different types of QWs to which such calculations were applied.  Thus, we recall here some basics facts only. 

 The analysis of the PL  from QWs with carriers   requires consideration of the LLs and this makes a difference with an analysis of the PL  from empty QWs. In the latter case and in the continuous-wave experiments, as in the present paper,  one can   assume that the relaxation of photoexcited minority carriers (holes in the case of QWs with a 2DEG)  is so fast that only  the PL from the lowest electron and hole levels occurs.  Then, one essentially avoids  $k$-dependent effects which result from the non-parabolicity and mixing of bands  at the band's extrema (we do not go here into $k$-dependent effects related to the interfaces~\cite{GBihlmayer_2015}).  

Within  the \kp (or \Pp - in non-zero $B$) theory one can approximately determine  the band structure taking into account $k$-dependent terms for given band and  the interaction between bands. The Luttinger model is   applied to the valence band, either to the four-fold degenerate $\Gamma_8^v$ band alone or to the $\Gamma_8^v$ together with  the spin-orbit-split $\Gamma_7^v$ bands~\cite{UEkenberg_1984,DABroido_1985,MKubisa_2003}. On the other hand, considering the  coupling of the CB and the VB in semiconductors with not-too-small band gap, the authors are rather concentrated on the influence of such a coupling on the CB without considering the VB.  In the present paper we apply this approach: the $\Gamma_6^c$ CB is treated with taking into account mixing of the bands while the $\Gamma_8^v$ VB is treated separately. However, there are authors which take into account interaction of bands in a symmetric way, determining the wave functions and the energies  in both the conduction and valence bands~\cite{WPotz_1985,FAncilotto_1988, VLopezRichard_2002}.

The effective $g$-factor, $g^*$, of electrons and holes is one of basic characteristics of carriers in semiconductors and indispensable in interpretation of optical spectra at $B$. 
The literature concerning the $g^*$ in CdTe and \cdteQW\, QWs is quite large.  A temperature dependence of $g^*$ in bulk CdTe was studied by EPR between 4 and 66 K and the results were interpreted within a 3LM derived by L. Roth~\cite{LMRoth_1959}.  A value of  $g^* = -1.65$ was obtained by the spin quantum beats technique~\cite{MOestreich_1996} in bulk CdTe~\cite{APHeberle_1994}. The same technique was applied to \cdteQW\, QWs where it was shown that the $g^*$ decreases with the well's width~\cite{QXZhao_1996}. For a QW width of 21~nm (i.e., practically the same as 20~nm considered in the present paper and a similar Mg content of 25\%) a value of -1.56 was obtained, which precisely coincides with the results of our calculations, as it will be shown further on. This is an argument supporting our conviction of a sufficient precision of the calculations of the CB wave functions within the proposed theoretical approach.

Electron's and hole's effective $g$ factors in \cdteQW\, QWs were also measured by a spin-flip Raman techniquein Ref.~\cite{AASirenko_1997}. In that extended work, electron's $g^*$-factors for $B$ parallel and perpendicular to the  QW plane were determined for QWs of different thickness.  An anisotropy of the hole's $g^*$-factor was also determined by a spin-echo technique~\cite{SVPoltavtsev_2020}. The effective $g$-factor was also measured in the presence of a 2DEG and a model was developed which took into account renormalization of the spin-orbit coupling resulting from the electron-electron interaction~\cite{EAZhukov_2020}.

The paper is organized as the following. Section~\ref{Experimental} describes the samples and the experimental system. In Section~\ref{Theory} we present a theoretical description of the valence and conduction bands which will be subsequently used in analyzing the data. Section~\ref{Results} contains presentation of the experimental data and its analysis. Finally, we conclude the paper in Section~\ref{Conclusions}. The appendix contains a short description of the wave functions in the CB resulting from mixing of bands. 

\section{Samples and the experimental set-up}{\label{Experimental}}

The samples used in the experiment are grown with the molecular beam epitaxy (MBE). Semi-insulating GaAs wafers with MBE-grown buffer layers are used as substrates. The active part of samples consists of modulation-doped CdTe quantum wells (QWs) with \cmt barriers. We study three samples: one of them contains only a single quantum well (SQW) and two others are  multiple quantum wells  (MQW) containing ten nominally identical (within a given sample) QWs. The samples' parameters are summarized in Table~\ref{Samples}. The width of iodine-doped layer is the same in all three  samples (12 monolayers, i.e., approximately 4~nm) and the width of the QWs  is always 20~nm.  The doped layer is separated from the well by a \cmt  spacer, the thickness of which influences the electron concentration in the well (it is higher for a 10~nm-thick spacer than a 20~nm-thick one). In the  MQW samples, neighboring QWs are separated by an undoped \cmt layer  the thickness of which depends on the spacer thickness, so that the length of the period is always 74 nm. The cap layers are composed of a 20~nm-thick  undoped~Cd$_{0.7}$Mg$_{0.3}\!$Te, a 1~nm-thick iodine-doped \cmt   to compensate a  surface charge   and a 10~nm-thick undoped  Cd$_{0.7}$Mg$_{0.3}\!$Te. 

\begin{table}[hbt!]
\centering
\caption{Parameters of \cdteQW\, QWs}
\label{Samples}
\begin{tabular}{|c|c|c|}
\hline
Sample &   Spacer [nm] &  Undoped layer [nm]    \\ \hline
SQW & 20 &   30   \\ \hline
MQW A &  20 &   30    \\ \hline
MQW B &   10 &  40   \\ \hline
\end{tabular}%
\end{table}

The samples are mounted in an insert and placed in the center of a 9~T superconducting magnet. They are cooled  to 1.8~K or 4.2~K.    The photoluminescence is excited with a 514~nm light from an Ar$^+$ laser. Spectra are measured with an analyzer of the circular $\upsigma^+$ and $\upsigma^-$ polarizations and a spectrometer with a CCD camera. Subsequent spectra are registered during 9~s each while the magnetic field is slowly  (0.00553 T/s) swept from 0 to 9~T. This allows us to register spectra every 0.05 T and it is  verified that the uncertainty in $B$  does not influence interpretation of spectra. Time-resolved PL measurements are carried out at $T$ = 5 K and zero $B$. In this case, the PL is excited with a 470~nm laser line (the second harmonics of a Ti:Spphire laser) and PL spectra are registered with a streak camera.

\section{The theoretical model}{\label{Theory}}
\subsection{The conduction band}

Since CdTe and Cd$_{0.7}$Mg$_{0.3}$Te layers forming the wells and   barriers, respectively, belong to medium-gap  materials, one should use for their description   a multiband formalism of the {\kp} theory. Specifically, we use a  three-level  model (3LM) of  the {\kp} theory. We consider the structure extending in the plane perpendicular to the growth direction $z$, so   the energy gaps and other band parameters are functions of $z$.  The system studied is considered invariant in the plane perpendicular to $z$. The model takes into account eight bands (including spin)  $\Gamma^c_6$, $\Gamma^v_8$   and $\Gamma^v_7$ levels at the center of the Brillouin zone; the band edge energies $E_{l}$ are $0$, $E_0$ and $G_0 = E_0 + \Delta_0$, respectively.  The distant (upper and lower) levels are treated as a perturbation and the resulting bands are spherical but nonparabolic.

 For the presence of a magnetic field, the {\kp} theory becomes the {\Pp} theory, with $ {\bf P} = {\bf p}+e{\bf A}$ replacing the momentum operator ${\bf p}$.  For the magnetic field $\textbf{B} || \textbf{z}$, we choose the Landau gauge for the vector potential: $\textbf{A} = [-By, 0, 0]$. 
 The complete eigenfunction in the above formalism for $n^{\mbox{\small{th}}}$ Landau level in  $i^{\mbox{\small{th}}}$ band is
\begin{equation}
\Psi_n^i =\sum_l f_{n+s(l)}^i c_{n+s(l)}^i u_l\;\;,
\end{equation}
where
\begin{align} \label{EnvelopeFunction}
	f_{n+s(l)}^i = \exp(ik_xx)\phi_{n+s(l)}[(y-y_0)/L]\chi_l^i(z) = \\ \nonumber
= F_{n+s(l)}^i\phi_{n+s(l)}.
\end{align}
Here $\phi_{n+s(l)}$ is the  harmonic oscillator function with the index $n+s(l)$, 
$y_0 = k_xL^2$ in which $L = (\hbar/eB)^{1/2}$ is the magnetic length and $c_{n+s(l)}^i$ are the participation coefficients of functions $u_l$ that satisfy the condition  $\sum_l |c_{n+s(l)}^i|^2 = 1$. The Bloch amplitudes $u_l$ are defined in Table~\ref{BlochFunctions}.

 The index $n + s(l)$ reflects  two aspects of the mixing of bands within the \Pp theory.  First, the LL with the index $n$ (i.e., the function $\Psi_n^i$)  contains the harmonic oscillator functions with indexes different from $n$. As an example, in the case of the CB, the $n_c^{\mbox{\small{th}}}$ LL contains harmonic oscillator functions with indexes $n_c, n_c-1, n_c+1$, so  $s = 0, \pm1$.  Second, the function $\phi_{n+s(l)}$ in the product with $u_l$ depends on $l$ which makes the number $s$ to be  $l$-dependent. A detailed description of the mixing in the CB is presented in the Appendix~\ref{KacmanZawadzki} and in the VB -- in the next subsection. 

With the above wave function, the \Pp Hamiltonian gets the form (see, e.g.,~\cite{PPfeffer_2003}; we abandon for a while the complex index of $f$): 
$$
\sum_l\left[\left( \frac{P^2}{2m_0}+E_{l}+V(z)-{\cal E}\right)\delta_{l'l} \right.
$$
\begin{equation}
\left.+\frac{1}{m_0}{\bf p}_{l'l}\cdot {\bf P} +\mu_{\rm B} {\bf B} \cdot {\boldsymbol{\sigma}}_{l'l} \right]f_l =0 \;\;,
\label{kpCB}
\end{equation}
where ${\cal E}$ is the energy   and ${\boldsymbol{\sigma}}_{l'l} = (1/\Omega)\langle u_{l'}|{\boldsymbol{\sigma}}|u_l\rangle$. Here ${\boldsymbol{\sigma}}$ are the Pauli spin matrices, $\Omega$ is the volume of the unit cell, $\mu_B = e\hbar/2m_0$ is the Bohr magneton, ${\mathbf{p}}_{l'l} = (1/\Omega)\langle u_{l'}|{ \mathbf{p}}|u_l\rangle$ are the interband matrix elements of momentum. The sum runs over all bands $l = 1, 2,...,8$ included in the model and $l'=1, 2,...,8$ runs over the same bands.
Within 3LM there exists the interband matrix element of momentum $P_0$ coupling the conduction band and the valence bands:
\begin{equation}
	P_0=\frac{-i\hbar}{m_0\Omega}<S|p_x|X> \; ,
\end{equation}
 and that of the spin-orbit interaction $\Delta_0$
 \begin{equation}
 	\Delta_0=\frac{-3i\hbar}{4m_0^2 c^2}<X|[{\bm \nabla} V_0,{\bf p}]_y|Z>.
 \end{equation}
 
\begin{widetext}

\begin{table}
	\caption{The periodic parts of Bloch functions ($u_l$) and corresponding energy levels ($E_l$) as used in the three-level {\kp} model. The total angular momentum $j$ and its $z$ component $j_z$ are also indicated.	}
	\begin{tabular}{|c|c|c|c|c|c|}
		\hline$j_z$&$u_l$&$j$&$E_l$&$u_l$&$j_z$\\\hline
		$\frac{1}{2}$&$u_1=iS\uparrow$&$\frac{1}{2}$ &0&$u_2=iS\downarrow$&-$\frac{1}{2}$\\  
		$\frac{1}{2}$&$u_3=\sqrt{1/3}R_+\downarrow$-$\sqrt{2/3}Z\uparrow$&$\frac{3}{2}$&$E_0$&$u_4=\sqrt{1/3}R_-\uparrow$+$\sqrt{2/3}Z\downarrow$&-$\frac{1}{2}$\\
		$\frac{3}{2}$&$u_5=R_+\uparrow$&$\frac{3}{2}$&$E_0$&$u_6=-R_-\downarrow$&-$\frac{3}{2}$\\
		$\frac{1}{2}$	&$u_7=\sqrt{2/3}R_+\downarrow$+$\sqrt{1/3}Z\uparrow$&$\frac{1}{2}$ &$G_0$&$u_8=\sqrt{2/3}R_-\uparrow$-$\sqrt{1/3}Z\downarrow$&-$\frac{1}{2}$\\\hline
	\end{tabular}
\label{BlochFunctions}
\end{table}

\end{widetext}

\begin{figure} 
\centering
\includegraphics[width=\linewidth] {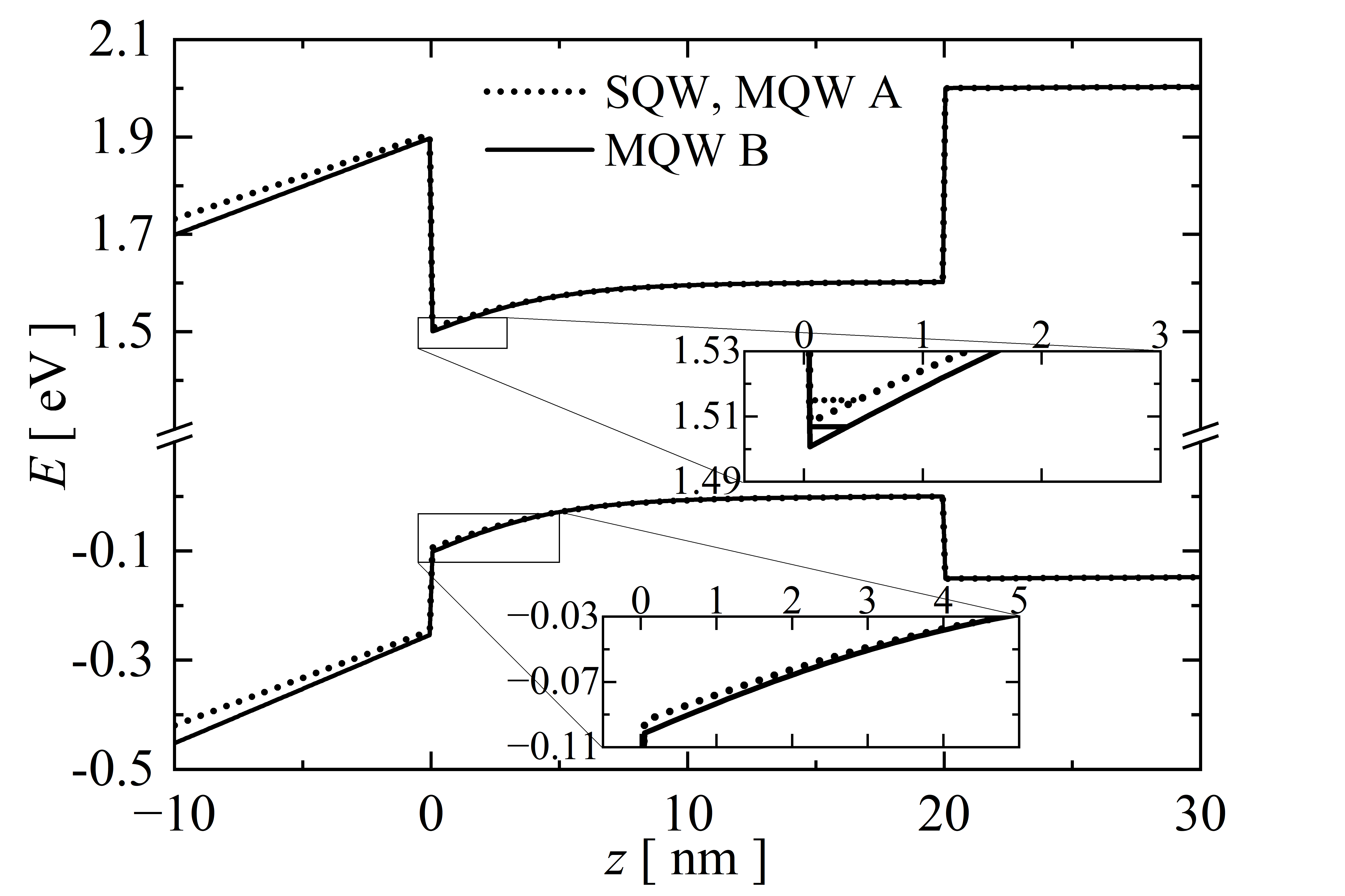}
\caption{\label{Potential} The self-consistent potential of the \cdteQW\,  QWs with parameters given in Table~\ref{Samples} (the growth direction is from right to left). The insets present enlarged parts of the main plots to show a difference in the potential for two different QWs.  The upper inset shows also position of the ground state in the SQW and  MQW A   samples at 6.258~meV and in the MQW B one at 6.175~meV above the well's lowest energy (the short horizontal dotted and solid line, respectively).  }
\end{figure}

The   potential $V(z)$ results from self-consistent Schr\"{o}dinger - Poisson calculations and is presented in Fig.~\ref{Potential}.  In these calculations, strain effects were not taken into account~\cite{JCibert_1995,BKuhn_1993}. The shape of the calculated potential, together with technologically known geometrical parameters of the structure and the doping level gives the electron concentration in the QW. The applied numerical procedure gave the values of the electron concentration equal to that obtained from transport measurements (not shown in the present paper). Since the  QWs are non-interacting because of a large separation, calculations of the potential for the  MQW samples was carried out in the same way as for the SQW sample. The same function $V(z)$ was used, for a given sample, to carry out both the conduction and valence bands calculations, described in this and the next subsection, respectively. 
The calculated potential $V(z)$ describes the CB offsets. The VB offsets are automatically determined by corresponding energy gaps.

The solution of Eq.~\ref{kpCB}, which is an 8$\times$8 system of equations for eight envelope functions $f_l(\textbf{r})$,  yields a linear combination of both conduction and valence bands wavefunctions. 
Since we are interested here  in the eigenenergies
and eigenfunctions for the CB, we express the valence functions $f_3,...f_8$ by the conduction functions $f_1$ and $f_2$, the latter descibing the spin-up and
spin-down states, respectively.  After substituting the equations for l'=3,...8 to the equations for l'=1, 2, the effective Hamiltonian for $f_1$
and $f_2$ functions becomes

\begin{equation}
\hat{H}=\left[ \begin{array}{cc}
\hat{A}^+&\hat{K}\\
\hat{K}^\dagger&\hat{A}^-
\end {array}
 \right]\;\;,
\label{Matrix22}
\end{equation}
where
$$
\hat{A}^{\pm}=V(z)-\frac{\hbar^2}{2}\frac{\partial}{\partial z} \frac{1}{m^*( {\cal E},
z)}\frac{\partial}{\partial z}+
$$
\begin{equation}
+\frac{\hbar\omega_c^0(n_c+1/2)}{m^*( {\cal E}, z)/m_0} \pm\frac{\mu_BB}{2}g^*( {\cal E}, z)
\;\;,
\end{equation}
in which $\omega_c^0=eB/m_0$. 
 The off-diagonal matrix elements
$\hat{K}$ and $\hat{K}^\dagger$ in Eq.~\ref{Matrix22} appear due to inversion asymmetry of
the system along the growth direction $z$, see Fig.~\ref{Potential}, which results in an
additional Bychkov-Rashba spin splitting. These terms are omitted because they are negligible~\cite{PPfeffer_2003}, particularly at high $B$. The electron effective mass $m^*$ and the spin $g^*$-factor are given by:
\begin{equation}
\frac{m_0}{m^*({\cal E}, z)}=1+C-\frac{1}{3}E_{P_0}\left(\frac{2}{\tilde E_0}+ \frac{1}{\tilde G_0}\right) \;\;
\label{EffectiveMass}
\end{equation}
and
\begin{equation}
 g^*({\cal E}, z)=2+2C'+\frac{2}{3}E_{P_0}\left(\frac{1}{\tilde E_0}- \frac{1}{\tilde G_0}\right)\;\;,
\label{EffectiveG}
\end{equation}
where ${\tilde E_0} = E_0 - {\cal{E}}+V(z)$ and ${\tilde G_0} = G_0 - {\cal{E}}+V(z)$. Here $G_0 = E_0 + \Delta_0$, $E_{P_0}=P^2_02m_0/\hbar^2$ and $C$ and $C'$ are far-band contributions calculated from Eqs.~\ref{EffectiveMass} and \ref{EffectiveG} for ${\cal E}$ = 0 and $V(z)$ = 0, taking known values of other parameters enetering these equations.  The effective masses $m^*$ and the $g^*$ values depend on the band structure and consequently are different for wells and barriers. They also depend on the energy due to bands' nonparabolicity. Values of band-edge parameters are given in Table~\ref{Parameters}.

Next, the eigenenergy equations $\hat{A}^+ \chi^+={\cal{E}}^+_n \chi^+$ and $\hat{A}^- \chi^-={\cal{E}}^-_n \chi^-$ are solved separately along the $z$ direction for the two spin states using the boundary conditions
\begin{equation}
 \chi^{\pm}|_+ = \chi^{\pm}|_-,\;\;\;
 \left[\frac{1}{m^*}\frac{\partial \chi^{\pm}}{\partial z}\right]_+ = \left[\frac{1}{m^*}\frac{\partial \chi^{\pm}}{\partial z}\right]_-
\end{equation}
 at each interface. In addition one deals with the boundary conditions at $z = \pm\infty$, where the wave function must vanish.

To be consistent with previous publications (e.g., \cite{PPfeffer_2003} and formulas therein), in the CB calculations we put the bottom of the CB at zero energy which resulted in the negative values of the band gaps in Table~\ref{Parameters}. These negative values should be substituted to Eqs.~\ref{EffectiveMass} and \ref{EffectiveG}. 

\begin{table}
\caption{Band parameters of barriers and wells, as used in the
calculations, $C$ and $C'$ are far-band contributions to the band-edge values.
Interband matrix element of momentum $P_0$ is taken to be independent of $z$: $E_{P_0}$ = 21.07 eV. Parameters of CdTe were taken from Ref.~\cite{WinklerBook_2003}, these for \cdteQW\, QWs were from a linear interpolation between parameters of CdTe and of MgTe from~\cite{Handbook_2004}. The electron $g^*$ for \cmt  was taken from an experimental data from Ref.~\cite{AASirenko_1997}.   }
{\label{Parameters}}
\begin{ruledtabular}

\begin{tabular}{ccc}
&Cd$_{0.7}$Mg$_{0.3}$Te&CdTe\\
\hline
 E$_0$(eV) & -2.143 & -1.6  \\
$\Delta_0$(eV)&-0.95&-0.95\\
m$^*_0$/m$_0$&0.118&0.093\\
g$^*_0$&-0.5&-1.66\\
C&-1.351&-1.781\\
C'&-0.2434&-0.1947\\
\end{tabular}
\end{ruledtabular}
\end{table}

\begin{figure}[H]
\centering
\includegraphics[width=\linewidth]{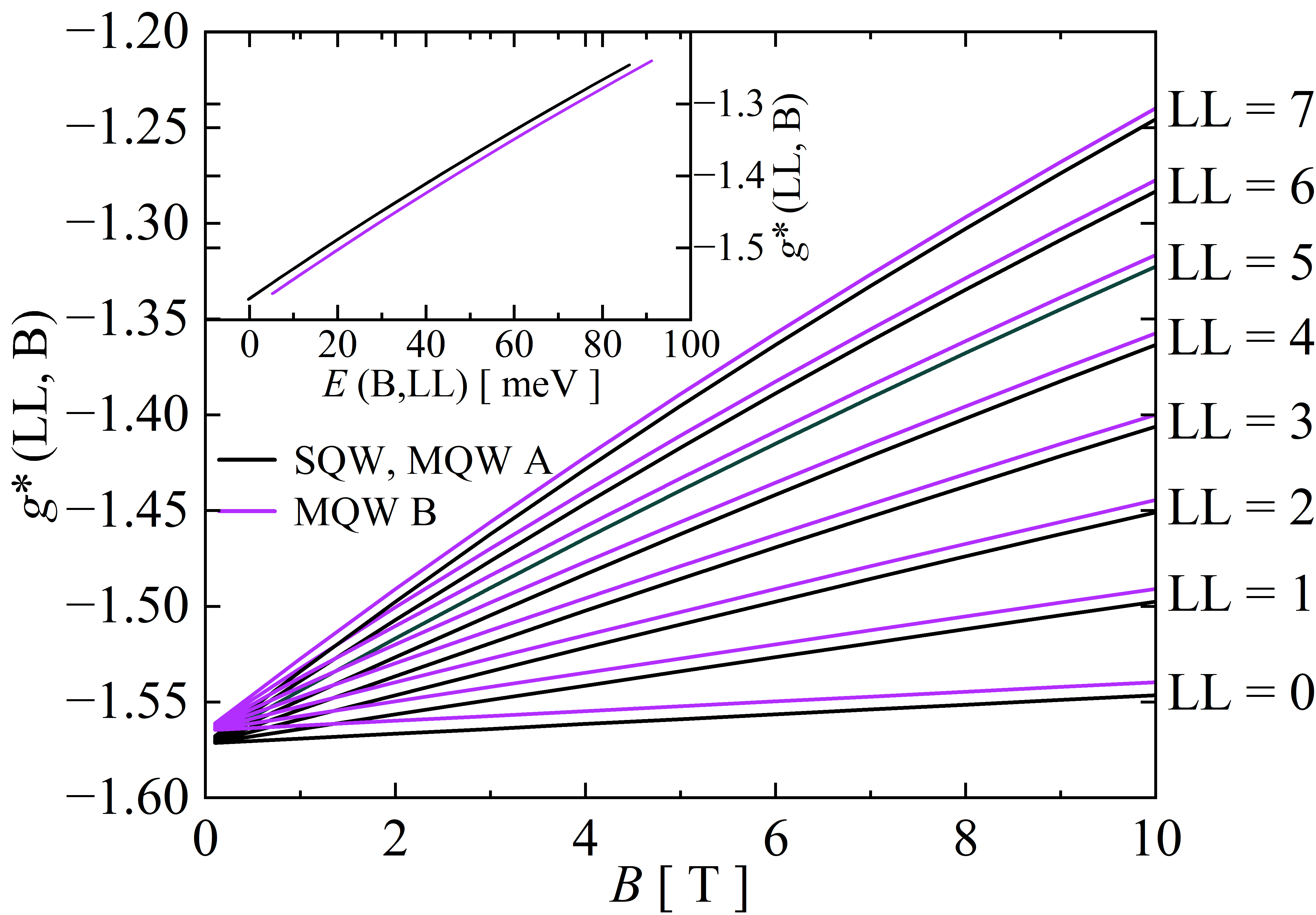}
\caption{The effective electron $g$-factor, $g^*$ ,for QWs with parameters given in Table~\ref{Samples}. }
\label{g_factor}
\end{figure}
In analyzing polarization - resolved PL spectra one has to use appropriate values of $g^*$. Results of calculations show that in the case of studied QWs, the electron $g^*$ decreases (in the absolute value) from the zero-field value of about -1.56, the change being stronger for higher LLs. The  value of $g^*$  depends on the energy only, independently on the number of the LL on which the electron resides. This is shown in the inset to Fig.~\ref{g_factor} where the description $E(B, LL)$ means that a given energy can be obtained on different LL and at different $B$.   Application of the results presented in Fig.~\ref{g_factor} gives a small correction to the energy in comparison with   a constant value of $g^*$. For example, in the case of $n_c$ = 7, assuming that $g^* \approx -1.25$ at 10~T gives a shift of the energy by only 0.17~meV in comparison with the zero-field value of about -1.56. For the lowest $n_c$ such  a shift is practically negligible but due to complication of spectra and many energetically close transitions, it would be not  reasonable  to totally neglect it.

In the case of the bulk material, the 3LM   can be solved analitically and leads to the general expressions for the wave functions of the $n^{\mbox{\small{th}}}_c$ LL~\cite{PKacman_1971}. It is found that the wave function $\Psi_{n_c}^c$ describing the $n^{\mbox{\small{th}}}_c$ LL is a linear combination of harmonic oscillator wave functions $\phi$ with index $n_c-1$, $n_c$ and $n_c+1$ (for $n_c=0$, the first term is omitted), as we mentioned above. The same mixing of oscillator wave functions occurs for a given LL  in the  case of QWs. The difference is that the presence of the self-consistent potential precludes obtaining analitical formulas for $\Psi_{n_c}^c$. The mixing of harmonic oscillator functions within a single LL influences  the optical selection rules which is disscussed in the next subsection.

A direct link, when the $n^{\mbox{\small{th}}}_c$ LL is composed of only the  $n^{\mbox{\small{th}}}_c$ oscillator function is no longer valid here. However, we will keep the name  $n^{\mbox{\small{th}}}_c$ LL, $n_c \ge 0$,    understanding that this level   tends to  $\phi_{n_c}$ at $B\rightarrow0$ when  terms with $\phi_{n_c\pm1}$ vanish. For the readers convenience, we quote in the Appendix~\ref{KacmanZawadzki} the form of the CB wave functions  obtained in~\cite{PKacman_1971} which shows the structure of mixed CB wave functions.

\subsection{The valence band }
Description of LLs in the VB is based on the Luttinger Hamiltonian~\cite{JMLuttinger_1955}  applied to the case of QWs at magnetic field. Here we follow the approach developed in Ref.~\cite{MKubisa_2003} for the case of a GaAs/(Ga, Al)As system. 
Details of the calculations are  presented in the full extent in~\cite{MKubisa_2003} and will not be repeated here. However, we will evoke some of the final results which are necessary in the data analysis, particularly from the point of view of the polarization selection rules of the optical transitions observed in the experiment.

The wave functions which describe LLs in the VB are  linear combinations of the form (keeping the numbering of the Bloch functions defined  in Table~\ref{BlochFunctions}):
\begin{equation}
\label{VBFunction}
\Psi_{n_v}^v = F_5^v\phi_{n_v-1}u_5 + F_3^v\phi_{n_v}u_3+F_4^v\phi_{n_v+1}u_4+F_6^v\phi_{n_v+2}u_6
\end{equation}
where $F_i^v$   are the envelopes to be determined. This is a particular case of a general expression given in Eq.~\ref{EnvelopeFunction}. Here we use the explicit form of the wave function $\Psi_{n_v}^v$ which allows us to write that the number $s(l)$ is equal in this case to -1, 0, 1, and 2 for $u_5, u_3, u_4$ and $u_6$, respectively, with an additional remark that the functions $\phi_{n_v + s(l)}$ vanish if their index is less than zero. 
 
%

\begin{figure}[H]
\centering
\includegraphics[width=\linewidth] {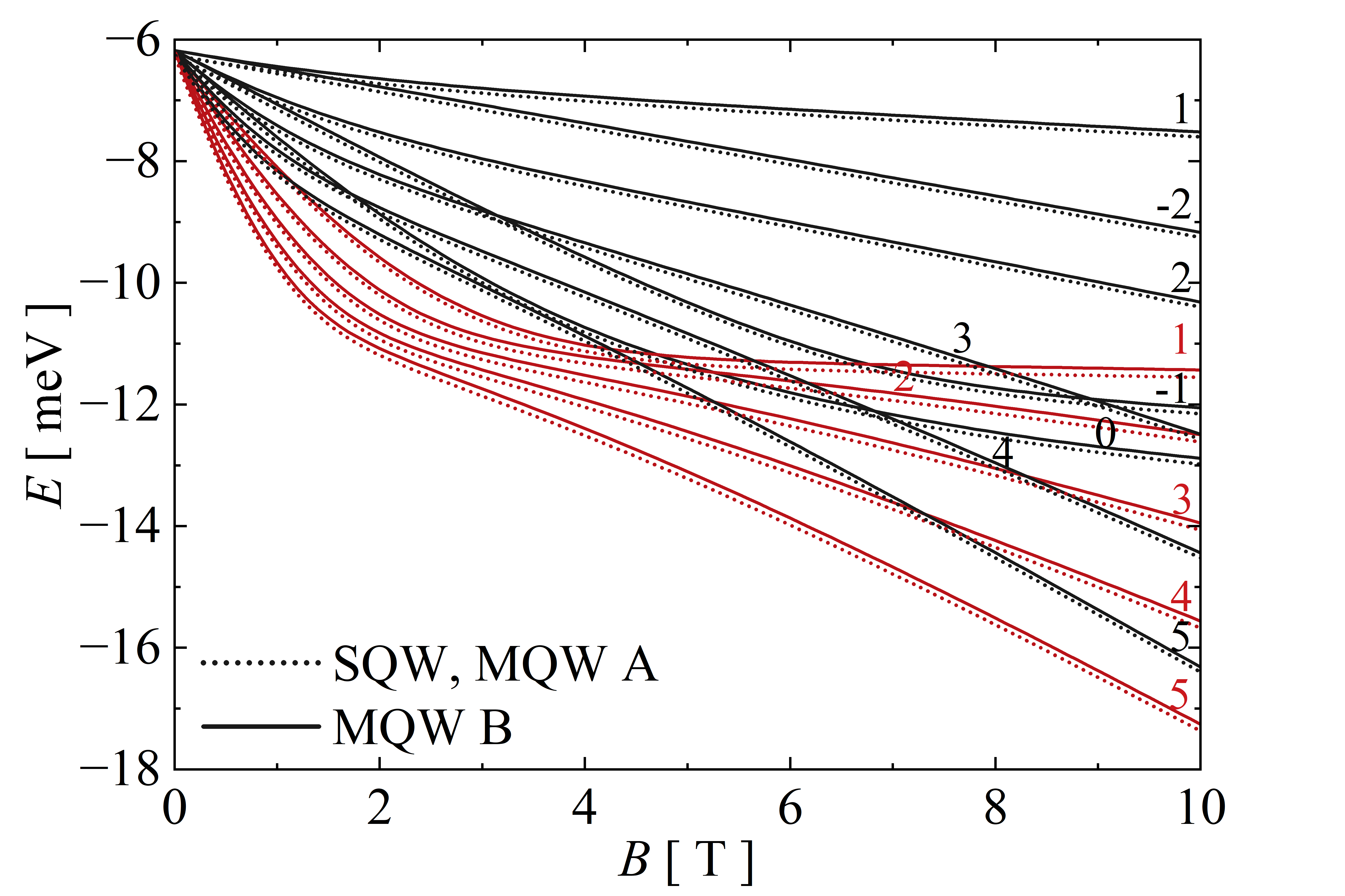}
\caption{\label{ValenceBand}Energy of LLs in the VB calculated for samples SQW and MQW A (dotted lines) and MQW B (solid lines). The numbers $n_v$ describing the  wave functions $\Phi_{n_v}$ are shown in the figure. }
\end{figure}

\begin{figure*}[ht!]
\centering
\includegraphics[width=\linewidth] {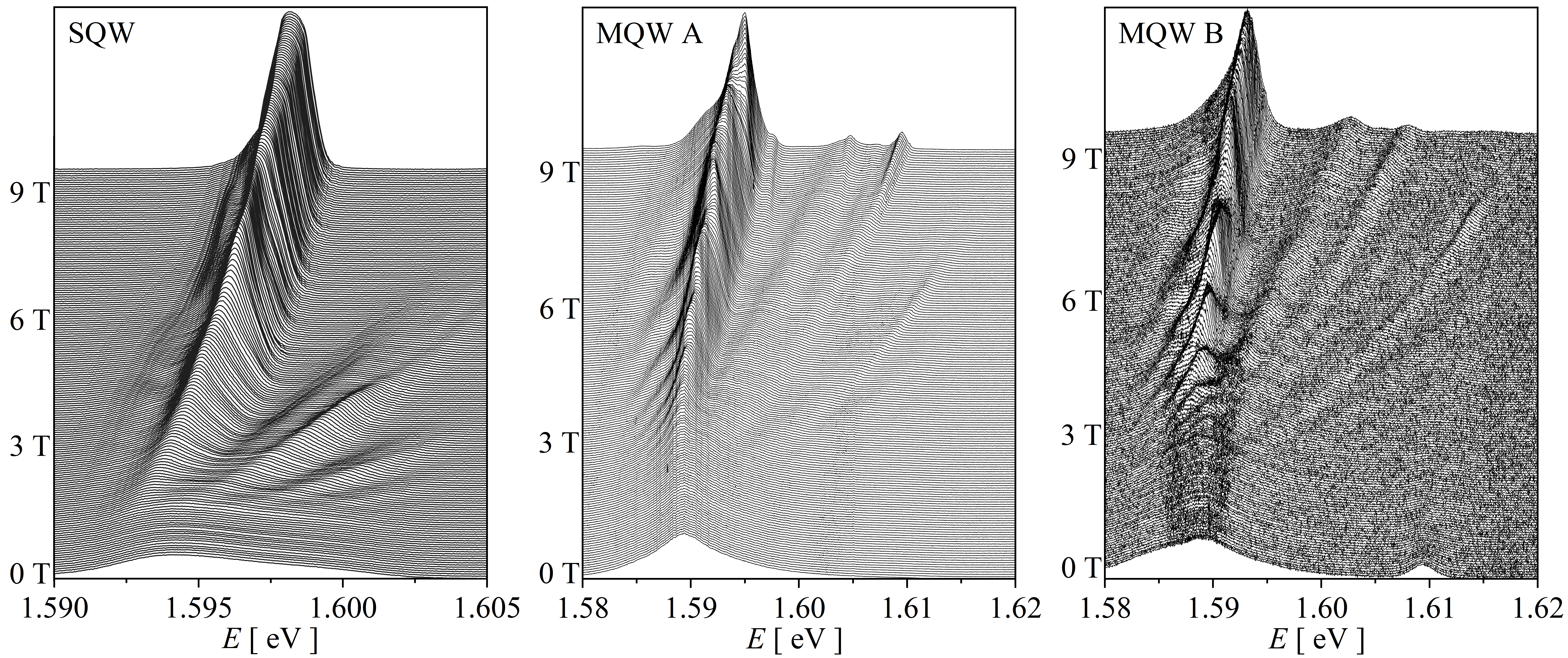}
\caption{\label{PL_spectra}PL spectra from the samples \cdteQW\, SQW, MQW A and MQW B (left to right) registered in $\sigma^+$ polarization. The PL signal from the MQW B sample was generally weaker than from the SQW and MQW A which is reflected by a higher level of the noise. }
\end{figure*}
 The calculations are carried out in the  axial approximation and their results are presented in Fig.~\ref{ValenceBand}. As one can see, except for $n_v$ equal to -2, -1, and 0, there are two sets of levels described by $n_v = 1, 2, 3...$, which are presented with red and black lines.  This reflects the fact that the final equation which allows one to determine the energy has two solutions for $n_v > 0$. As one can also notice, a linear dependence of LLs on $B$  is found only at $B$ up to about 1~T.  
As it was  in the case of the CB, we will keep naming  the LLs in the VB by the index  $n_v$ understanding that the  $n^{\mbox{\small{th}}}_v$ LL in the VB is composed of four oscillatory functions given by Eq.~\ref{VBFunction}. It is clear from Fig.~\ref{ValenceBand} that the ordering of the LLs depends on $B$ and has nothing to do with the simple (i.e., monotonic in $n_v$) ordering scheme found for a parabolic and isotropic band.

The theory presented in~\cite{MKubisa_2003} allows one to determine selection rules of optical transitions in the $\sigma^+$ and $\sigma^-$ polarizations. Under the assumption that the electron wave functions in the CB are of the form $\Psi_{n_c}^{c} = F_{n_c}^c\phi_{n_c}u_1\equiv \Phi^c_{n_c}\!\!\uparrow$ and $\Psi_{n_c}^{c} = F_{n_c}^c\phi_{n_c}u_2\equiv \Phi^c_{n_c}\!\!\downarrow$, so  there is no mixing of the wave functions in the CB (i.e., $s(l)=0$ for all LL in the CB in Eq.\ref{EnvelopeFunction}), one gets 
\begin{equation}
\begin{aligned} 
\Psi_{-2}^{v}&{}\; \longleftrightarrow\quad\, \upsigma^+\!:\:\:\Phi_{0}^c\!\!\downarrow\\[5pt]  
\Psi_{-1}^{v}&{}\; \longleftrightarrow\quad\, \upsigma^+\!: \:\:\Phi_{0}^c\!\!\uparrow,\;\, \Phi_{1}^c\!\!\downarrow\\[5pt] 
\Psi_0^{v}&{}\;\longleftrightarrow\left\{
\begin{array}{cll}
\upsigma^-\!:& \Phi_{0}^c\!\!\downarrow& \\[5pt]
\upsigma^+:& \Phi_{1}^c\!\!\uparrow,& \Phi_{2}^c\!\!\downarrow
\end{array}
\right.\\[5pt]  
\Psi_{n>0}^{v}&{}\;\longleftrightarrow\left\{
\begin{array}{ccl}
\upsigma^-\!:& \Phi_{n-1}^c\!\!\uparrow,& \Phi_{n}^c\!\!\downarrow\\[5pt]
\upsigma^+:& \Phi_{n+1}^c\!\!\uparrow,& \Phi_{n+2}^c\!\!\downarrow
\end{array}
\right.
 \end{aligned}
\label{SelectionRules}
\end{equation}
The spin of $\Psi_{n_v}^v$ is not indicated because these functions contain components with both spin directions, as can be seen from their  structure given by Eq.~\ref{VBFunction}.  In the following we will consider  a set of more general selectiun rules which appears to be necessary to describe our experimental data.

\section{Results and discussion}{\label{Results}}

\subsection{The correspondence of the theoretical calculations and experimental data}
Figure~\ref{PL_spectra} shows PL spectra measured in the $\sigma^+$ polarization. A very similar picture of the PL measured in $\sigma^-$ polarization is not presented because the degree of polarization is generally small in the samples studied, the energies of   transitions in~$\sigma^+$ and $\sigma^-$  differ only slightly and such small differences are hardly visible in the waterfall presentation. The spectra from the SQW sample cover a narrower  energy range (about 1.592 to 1.604 eV) than those of the  MQW samples (about 1.590 to 1.616~eV). The shapes of waterfalls presented in the three panels are generally very similar one to another, with the main peak growing in the intensity with the increase of $B$ and well-resolved structures at the high energy part of the spectra. The latter is attributed to transitions between LLs in the conduction and valence bands, as it will be discussed further on.
 The MQW B sample differs from the other two by a  PL from the second subband (around 1.61 eV) which is visible in this sample only. This is due to the narrower spacer than in SQW and MQW A which results in a higher electron concentration. 
\begin{figure}
    \centering
    \includegraphics[scale=0.33]{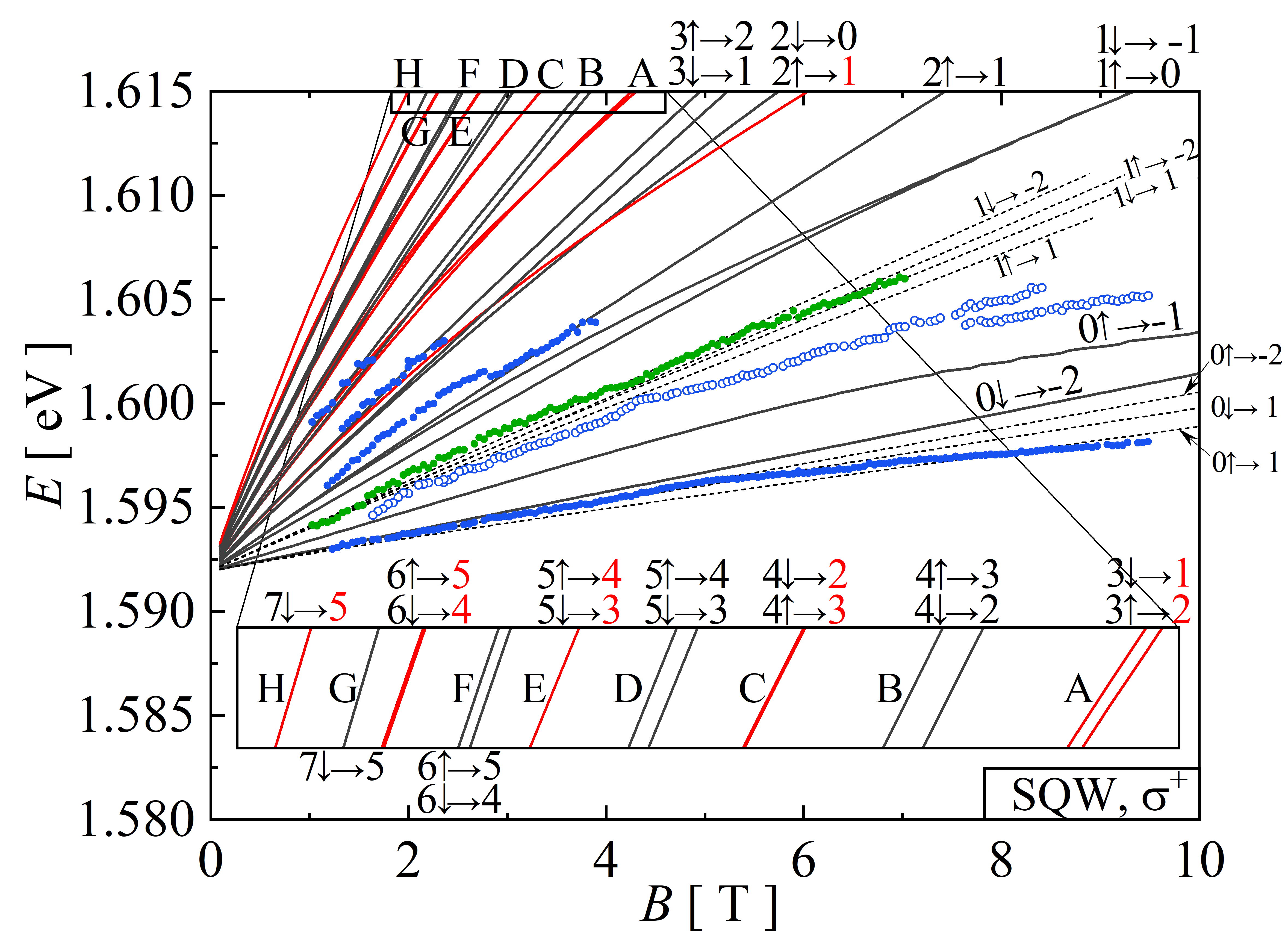} 
   \caption{ Positions of spectral features for the \cdteQW\, SQW sample in the $\sigma^+$  polarization (points). Solid lines are results of model calculations described in the text.  The labels on the top (and in the panel - enlarged) and right indicate  LLs and the electron spin in the CB (the first number) and the  LL in  the VB (the second number). Red and black colors correspond to LLs in the VB marked in red and black in Fig.~\ref{ValenceBand}. Green points - the ``missed'' transition. Open points mark transitions with the same energy observed in both polarizations (see Fig.~\ref{SQW_sigma-}). Dashed lines - transitions with relaxed selection rules given by Eq.~\ref{SelectionRules}.}
    \label{SQW_sigma+} 
  \end{figure}
  \begin{figure} 
    \centering
    \includegraphics[scale=0.33]{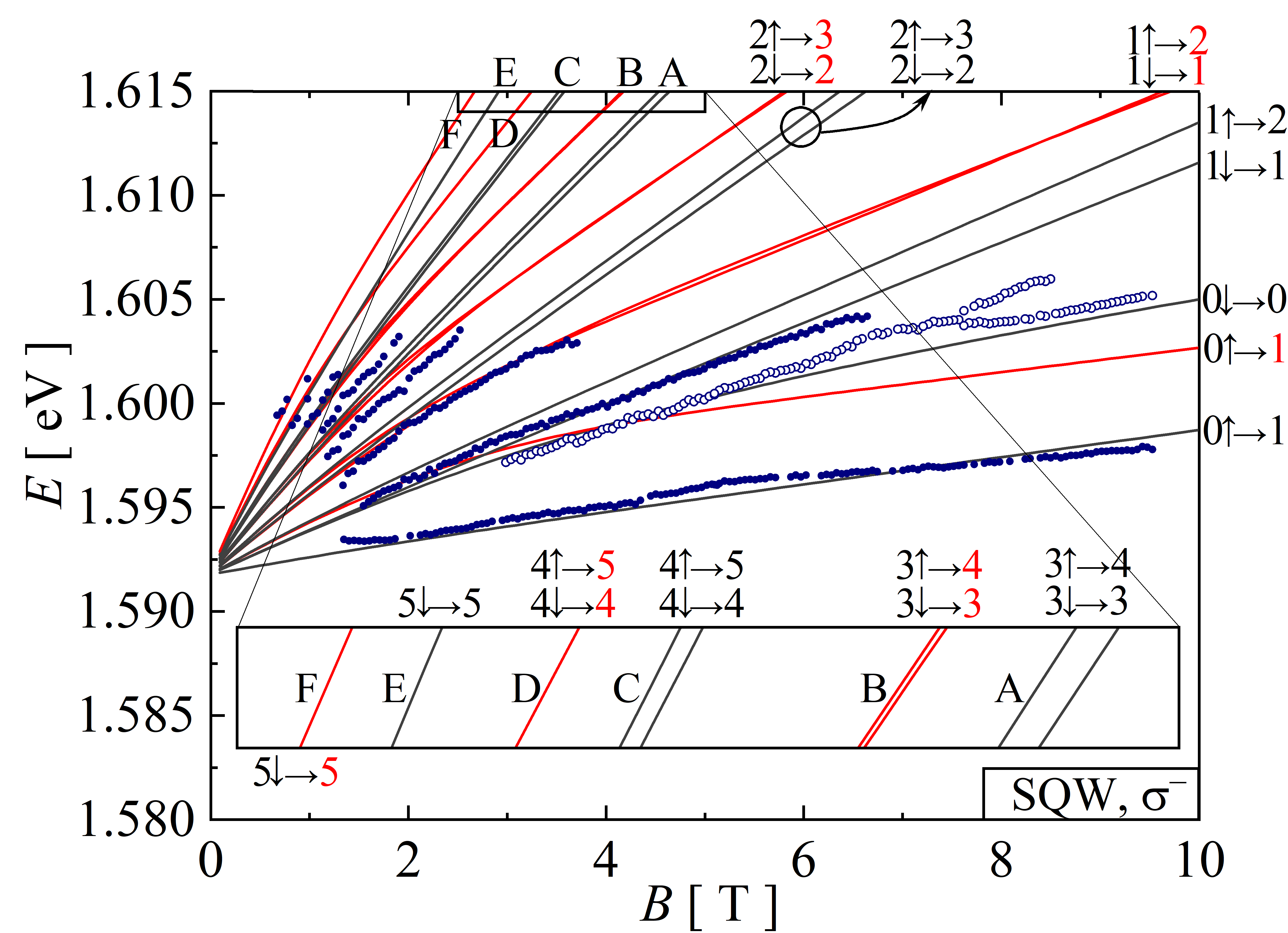} 
     \caption{  The same as in Fig.~\ref{SQW_sigma+} but in the $\sigma^-$ polarization.  }
    \label{SQW_sigma-} 
  \end{figure}

To compare the results of the theoretical calculations with experimental data, we select  such  pairs of   levels (one from the CB and the other from the VB) which correspond to transitions allowed by the selection rules presented above.  A free parameter  is the energy gap $E_0$, corrected for the confinement energies.  The value  of $E_0$ is adjusted by sliding against each other the two graphs - one with experimentally measured transition energies and the other - with theoretically obtained dependencies. Thus, the final relative position of the two graphs does not result from any numerical optimalization  procedure. On the other hand, we pay  attention to keep the   same  origin (i.e., the point a $B=0$)  of the sets of theoretical lines for both polarizations. Also, as one can observe in Fig.~\ref{SQW_sigma+} and \ref{SQW_sigma-}, there  are well-defined transistions from higher LL at $B$ less than about 4~T which help to achieve consistency between the experimental data and the theoretical description. 
 
We start the analysis by considering the results for the SQW sample. The position of peaks visible in spectra measured for the two polarizations are presented in Figs.~\ref{SQW_sigma+} and \ref{SQW_sigma-}, respectively. The theoretical analysis presented above refers to the recombination of  free electrons and free holes and is not applicable to  bound-to-bound or free-to-bound transitions. The structure of PL spectra measured is rather complicated and discrimination between free-to-free and other transitions is not easy, especially when practically the only tool to do so is a comparison of the data with theoretical (approximate, obviously) calculations. We note that there are transitions (marked with open points in Figs.~\ref{SQW_sigma+} and \ref{SQW_sigma-}) which are unpolarized. They also do not fit to any theoretically predicted transitions and most probably are not related to free-to-free transitions. 

The solid lines in Figs.~\ref{SQW_sigma+} and \ref{SQW_sigma-} describe the observed transitions quite well with the only exception of points marked with green symbols in Fig.~\ref{SQW_sigma+}.  As one can see in Fig.~\ref{SQW_sigma+}, these points fall within a ``gap'' between    $0\!\!\uparrow \rightarrow -1$ and $1\!\!\uparrow \rightarrow 0$ lines where no theoretically predicted transitions in the $\sigma^+$ polarization are present.

\subsection{The ``gap`` and  relaxation of the  selection rules}
Although the ``gap'' involves only one line of transitions (green points; we neglect transitions marked with open symbols) its appearance requires explanation because the theoretically missing transitions is apparently a partner $1\!\!\downarrow \rightarrow 1$ transition observed in  the $\sigma^-$ polarization and it is hard to understand why it is not theoretically predicted.
\begin{figure}[ht!]
\centering
\includegraphics[scale=0.5]{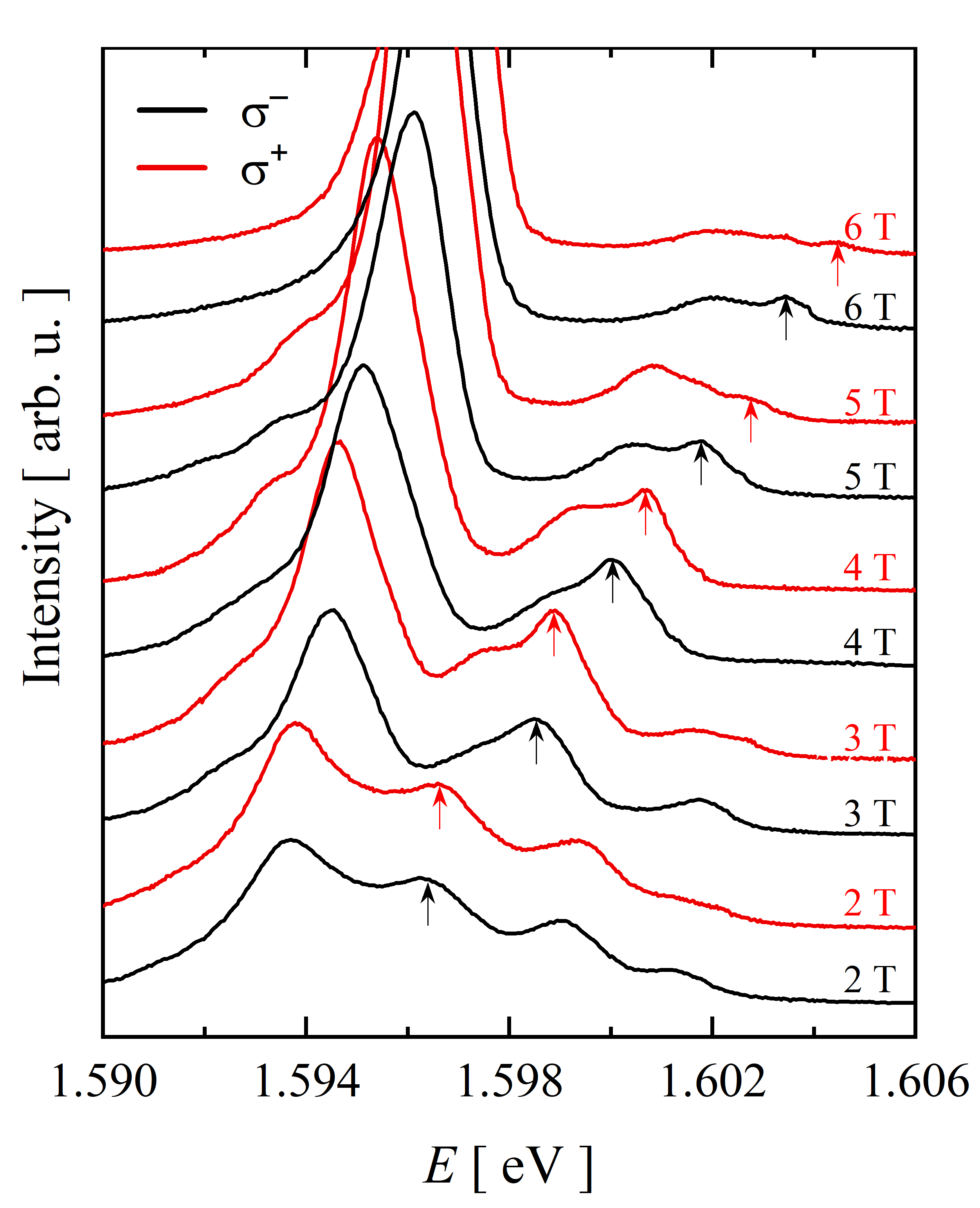}
\caption{PL spectra of the \cdteQW\, SQW sample showing the transitions (marked with arrows) tentatively ascribed as involving the $n=1$ LL in the CB. Red and black  curves - $\sigma^+$ and $\sigma^-$, respectively. }
\label{SQW_LL1}
\end{figure}
 Our reasoning is explained with the help of Fig.~\ref{SQW_LL1} which shows a set of spectra in the $\sigma^+$ (red) and $\sigma^-$ (black) polarizations.  The shape of the spectra at 2, 3 and 4 T with subsequent disappearing of the highest energy maximum  makes us to attribute these maxima to   the LLs in the CB with the number $n_c = 0, 1, 2, ...$. Such shapes of spectra are typically observed in the PL from a two-dimensional electron gas  and correspond to an increasing density of states at LLs as $B$ grows~\cite{KMeimberg_1997, JLusakowski_2011}.  Now, we concentrate only on the peaks marked with arrows which, according to the above  assumption, result from recombination of electrons from $n_c = 1$ LL in the CB. As it is shown in Fig.~\ref{SQW_LL1}, the separation of the corresponding maxima in $\sigma^+$ and $\sigma^-$ increase with $B$ which is a natural consequence of a spin splitting.


 To resolve the problem of  the ``gap'', we analyze more closely selection rules of free-to-free transitions in the system studied. Generally, we are looking for a way to enlarge the set of selection rules presented in Ref.~\cite{MKubisa_2003} and Eq.~\ref{SelectionRules}.
 The allowed transitions correspond to non-zero matrix elements $\langle\Psi_{n_v}^h|\boldsymbol{\epsilon}\cdot\boldsymbol{p}|\Psi_{n_c}^e\rangle$, where $\boldsymbol{\epsilon}$ describes the polarization of the  light. We are interested in circular polarizations given by  $\epsilon_x\boldsymbol{e}_x  + i \epsilon_y\boldsymbol{e}_y$ and $\epsilon_x\boldsymbol{e}_x  - i \epsilon_y\boldsymbol{e}_y$ corresponding to the $\sigma^+$ and $\sigma^-$ polarizations, respectively.  Thus, in the product $\boldsymbol{\epsilon}\cdot\boldsymbol{p} = \boldsymbol{\epsilon}^+ \cdot \boldsymbol{p}^- + \boldsymbol{\epsilon}^- \cdot \boldsymbol{p}^+ + \boldsymbol{e}_z \boldsymbol{p}_z$, where $\boldsymbol{p}^{\pm} = p_x\boldsymbol{e}_x  \pm i p_y\boldsymbol{e}_y$, only   two first terms  are of interest.

\begin{figure*}[ht!] 
  \begin{subfigure}{0.49\linewidth}
    \centering
    \includegraphics[width=\linewidth]{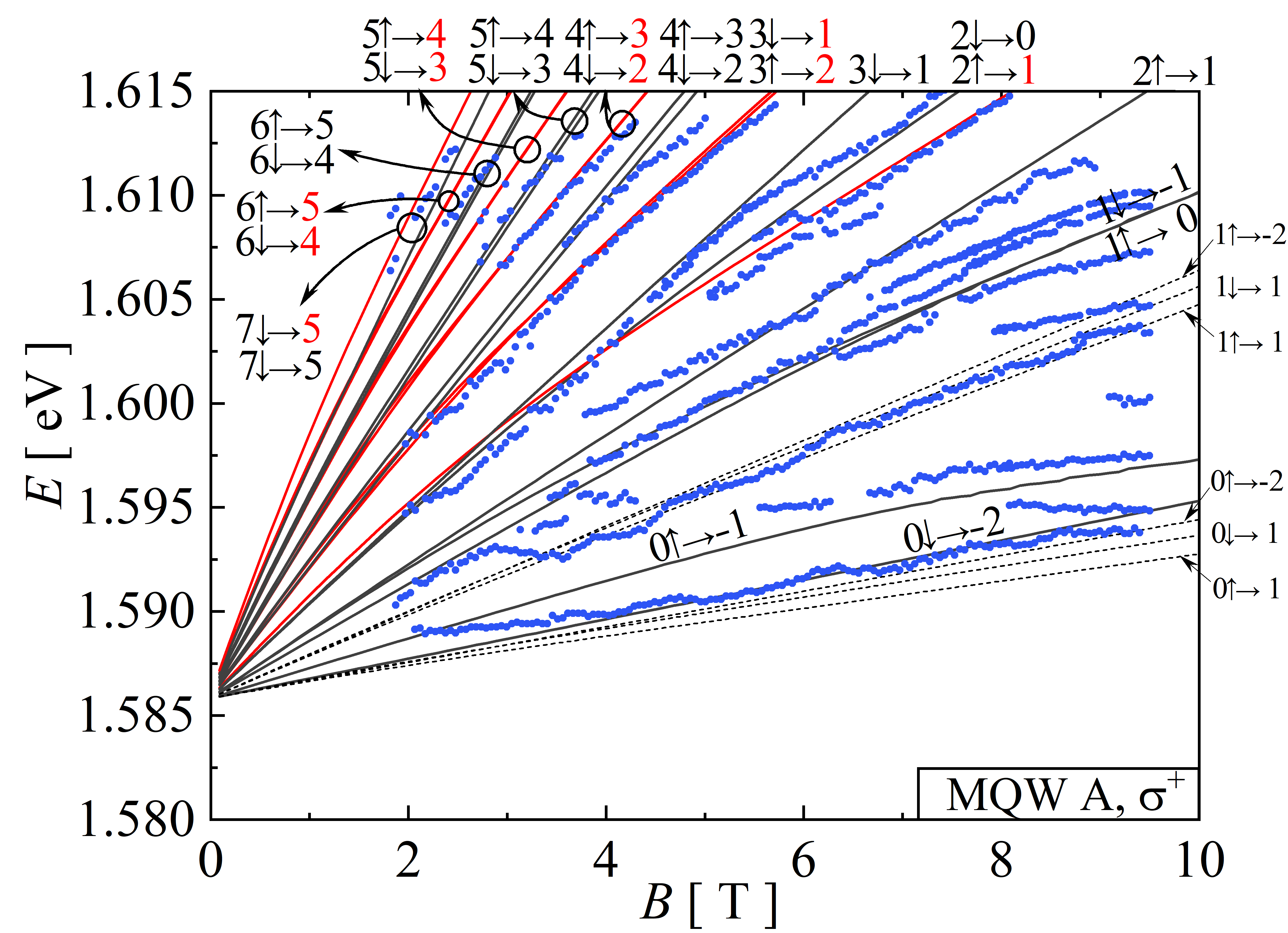} 
    \label{MQWA_sigma+} 
  \end{subfigure}
  \begin{subfigure}{0.49\linewidth}
    \centering
    \includegraphics[width=\linewidth]{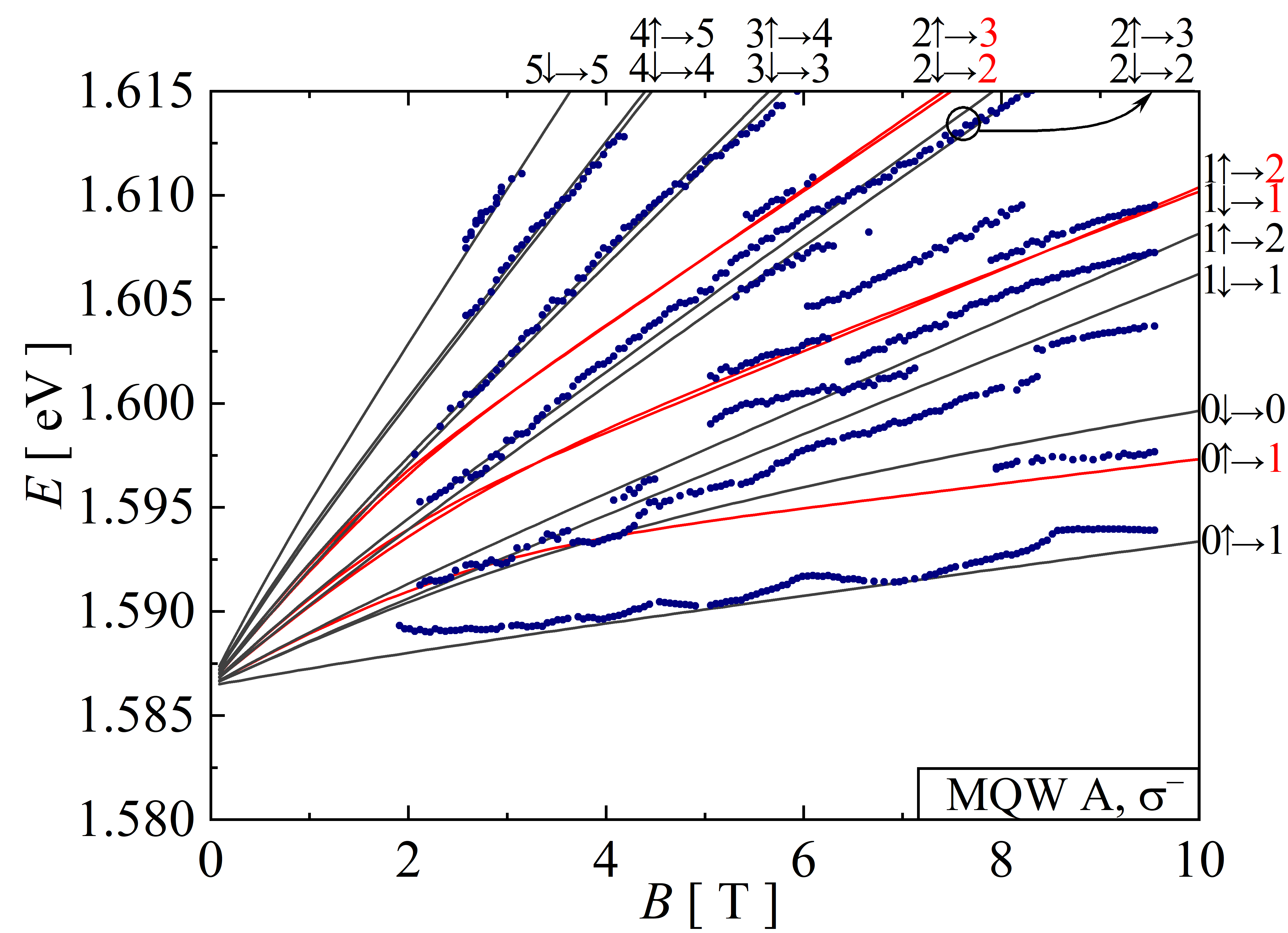} 
    \label{MQWA_sigma-} 
  \end{subfigure}

  \begin{subfigure}{0.49\linewidth}
    \centering
    \includegraphics[width=\linewidth]{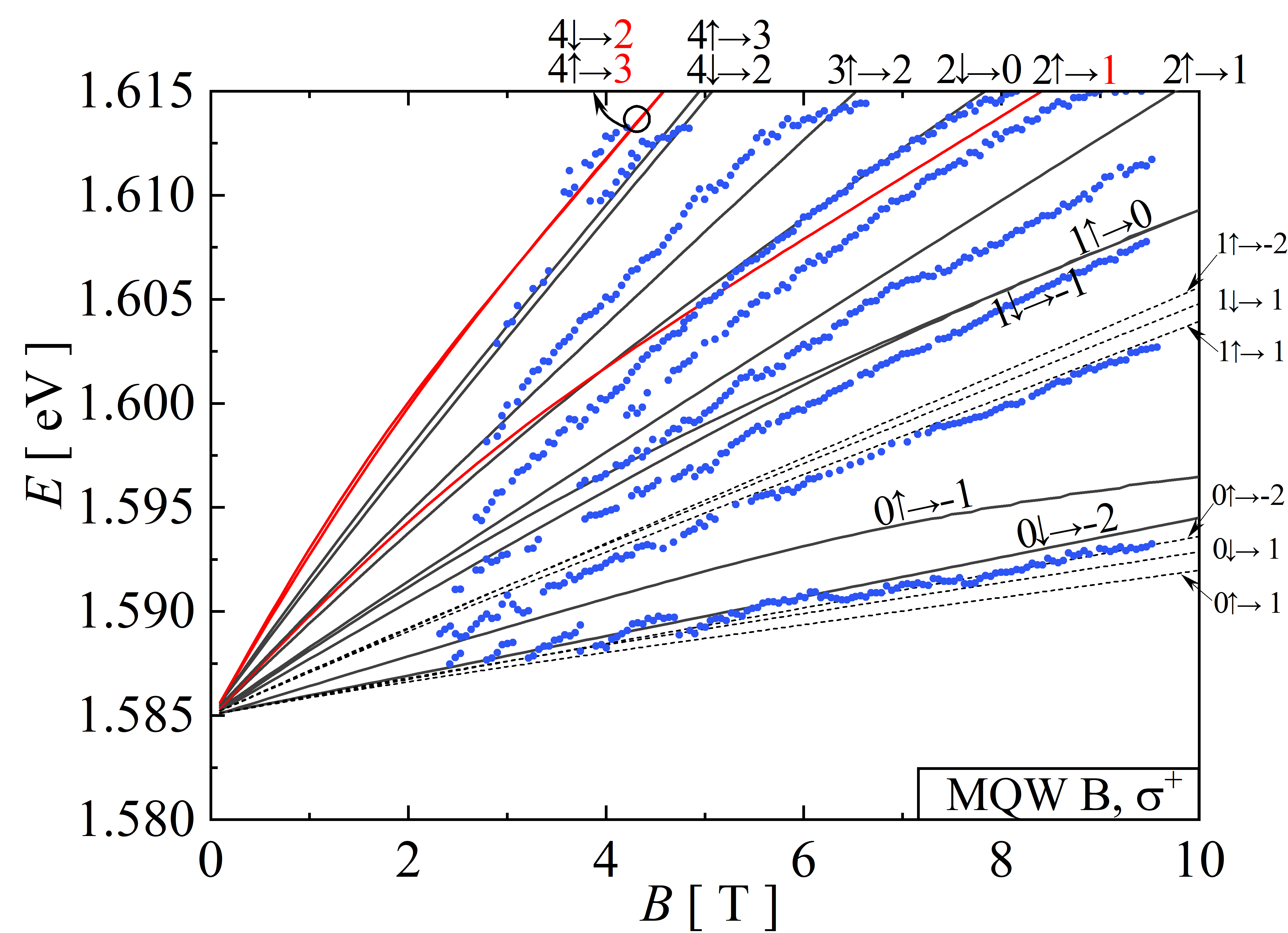} 
    \label{MQWB_sigma+} 
  \end{subfigure}
  \begin{subfigure}{0.49\linewidth}
    \centering
    \includegraphics[width=\linewidth]{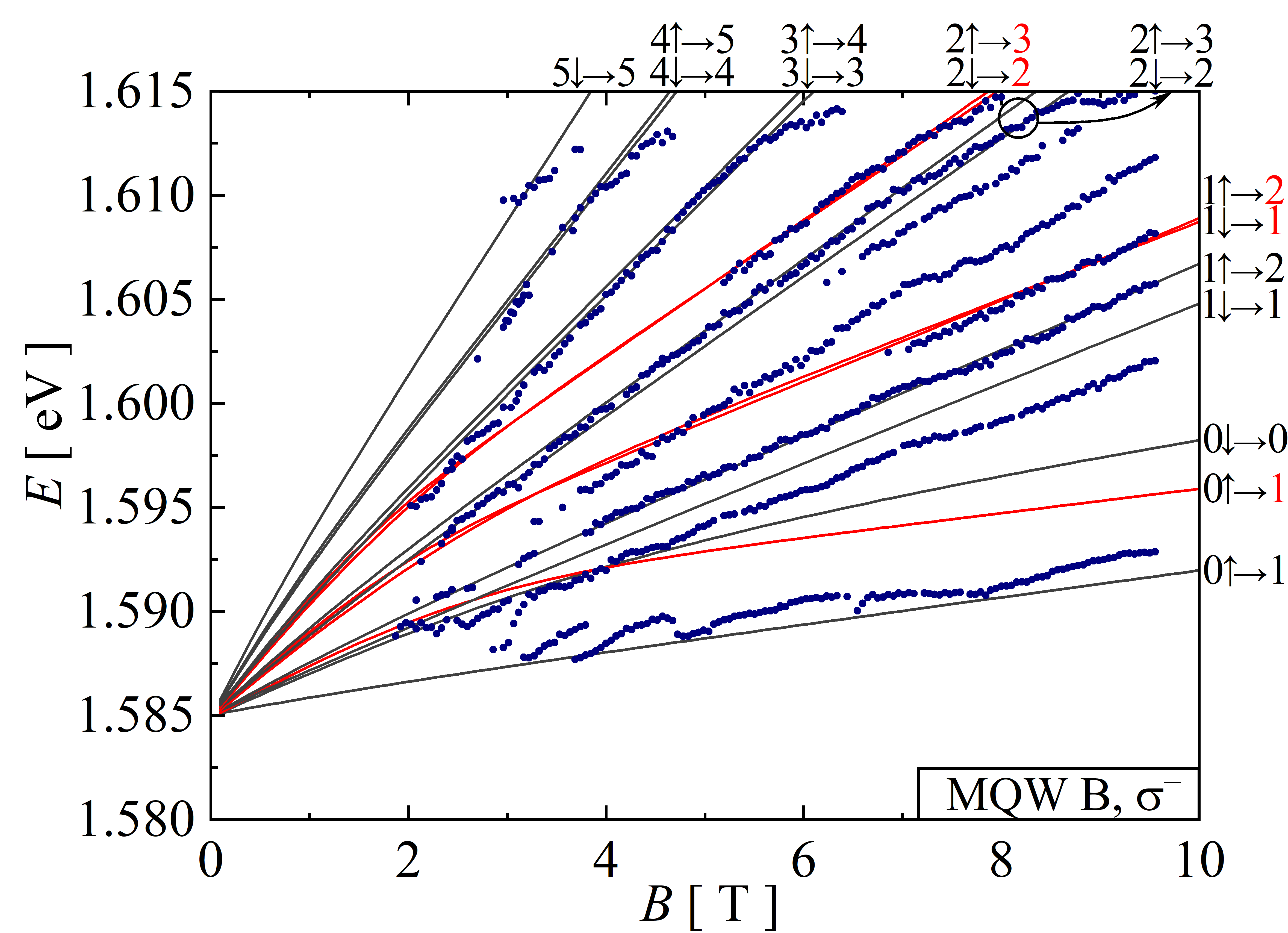} 
    \label{MQWB_sigma-} 
  \end{subfigure}
  \caption{Positions of PL peaks in data registered on the \cdteQW\, MQW A and MQW B samples in both polarizations, as indicated in figures. Other details of figures are the same as in Fig.~\ref{SQW_sigma+}.}
  \label{MQWData}
\end{figure*}

The general requirement of the selection rules leading in our case to allowed transitions  is that the matrix elements between the Bloch functions $u_i$ must be non-zero   (this involves also the requirement that the spin projection in the two involved functions $u_i$ is the same) and that the harmonic oscillator functions in the conduction and valence bands have the same index. The selection rules which were derived in Ref.~\cite{MKubisa_2003} and presented in Eq.~\ref{SelectionRules}  considered matrix elements between  the hole levels $\Psi_{n_v}^{v}$ given above and  CB functions of the $S$ symmetry.  As we showed above, interaction of the conduction and valence bands   makes the CB wave functions a mixture of $S$ and $X\pm iY$ functions with appropriate directions of the spin, as derived in Ref.~\cite{PKacman_1971}. However, this mixture does not generate new transitions because new   matrix elements would be proportional to $\boldsymbol{\epsilon}^{\pm}\cdot\langle X\pm iY |\boldsymbol{p}^{\mp}|X \pm iY\rangle$ and equal to zero because they are antisymmetric. 

Another possibility  was considered in Refs.~\cite{ KRyczko_2004, JJadczak_2012} where   an influence of the cubic terms in the  valence-band Hamiltonian on the selection rules was analyzed and it was  found   that these terms mix the VB wave functions with indexes $n_v$ and $n_{v\pm 4}$. This led to new theoretically allowed transitions which was necessary to explain experimental data. Also, including these terms as a perturbation to the axial Hamiltonian allowed to modify slightly the energy of the holes' levels and to achieve a more accurate description of the data. This, however, does not solve our problem because assuming that $n_c$ in $\Phi^e_{n_c}$ is equal to 1, we would potentially add transitions from LLs in the VB described by $\Psi_{n_v}^h$ with $n_v$ equal to at least 4. The energy of such transition would be much higher than awaited (see Fig.~\ref{ValenceBand}; we are expecting transition to the lowest VB  energy levels, i.e., $\Phi_1^v$ or $\Phi_{-2}^v$). 

Let us note that the  theoretical approaches to the CB and the VB presented above are not equivalent. In the case of the CB, to go beyond a simplified parabolic and one-band approach, we introduce an interaction of the CB $\Gamma_6^c$ with $\Gamma_8^v$ and $\Gamma_7^v$ in the VB. On the other hand, when considering the VB, only mixing of the four $\Gamma_8^v$ levels is considered, without any interaction with the CB. Trying to describe the bands and the interband optical transitions as accurately as possible, we are still bound by  an approximate description. 
  
We  propose that the ``missing'' transition becomes allowed due to   admixture of $S$ - type $\Gamma_6^c$ wave functions, with both direction of the spin,  to the $\Gamma_8^v$ band which accompanies an admixtures of $\Gamma_8^v$ wave functions to the $\Gamma_6^c$ CB (the latter being described by the 3LM). Such a mutual mixing is natural in the frame of \kp models and seems to be the simplest explanation of the results observed in this work. The strength of the mixing requires additional calculations  unifying both theoretical approaches described above which is beyond the scope of the present paper and left for further studies.  

The matrix element which would lead to  the ``missing'' transition are of the form  
$
 \boldsymbol{\epsilon}^+ \cdot \langle S\uparrow  |\boldsymbol{p}^-|(X- iY) \uparrow \rangle
$
or 
 $
 \boldsymbol{\epsilon}^+ \cdot \langle S\downarrow  |\boldsymbol{p}^-|(X- iY) \downarrow \rangle
$
  where the bra corresponds to the VB  and the ket - to the CB. Transitions described by these matrix elements are possible due to terms proportional to $\alpha_4$ and $\beta_4$ in Eqs.~\ref{Psi-} and~\ref{Psi+}.  Energy of all possible transitions from the $n_c = 1$ level to appropriate levels in the VB, i.e., with $n_v=1$ or $n_v=-2$ are shown with the dashed lines in Fig.~\ref{SQW_sigma+}. Although it is not possible to precisely determine which of these transitions is the 
 very one looked for, one can notice that their energy falls precisely in the region of the ``gap'' and agreement with the experimental data is quite satisfactory. 
 
We also apply  the relaxed selection rules to the transition  of the lowest energy which in Fig.~\ref{SQW_sigma-} is described as $0\!\!\uparrow \rightarrow 1$ and in Fig.~\ref{SQW_sigma+} at low $B$ as $0\!\! \downarrow \rightarrow  -2$. Similarly to the transitions from $n_c=1$ level, we plot in Fig.~\ref{SQW_sigma+} with dashed lines all transitions from the CB $n_c = 0$ LL to the highest $n_v = -2, 1$ LL in the VB. The energy of transitions vs. $B$ from $n_c = 1$ and $n_c=0$   seems to pass through different energetically possible levels which could suggest that observed transitions involve pairs of levels which change with $B$. Without drawing the definitive conclusion, we notice that this could be justified by  $B$-dependence of the coefficients $\alpha$ and $\beta$ in Eqs.~\ref{Psi-} and \ref{Psi+}.

\subsection{Multiple quantum wells}

Luminescence data obtained on the \cdteQW\,  MQW samples, each containing ten QWs, was treated in the same way as the SQW data was. The results are presented in Fig.~\ref{MQWData}. The structure of the PL is much richer than in the case of the SQW.    This is reflected by an essentially wider range of the energy of transitions visible in the PL and also by  transitions at the low-energy wing of the main peak (see Fig.~\ref{PL_spectra}). These structures are phonon-assisted transitions between LLs, they were discussed (in the case of the sample MQW B) in Ref.~\cite{WSolarska_2023} and will be not considered here; they are not presented in Fig.~\ref{MQWData}.

The solid lines in panels (a) and (b) in Fig.~\ref{MQWData} are the same as in Figs.~\ref{SQW_sigma+} and \ref{SQW_sigma-} because the QWs in these two samples are nominally identical. A much wider range of the energy covered by the PL spectra in  Fig.~\ref{MQWData} must be related to a higher electron concentration in the case of MQWs although the conditions of doping of all QWs (both in the SQW and MQW samples) were identical. The difference results from a certain technological drawback of  iodine doping which is the difficulty in eliminating   iodine from the MBE machine once iodine effussion cell has been opened for the first time. This results in ``drawing''   iodine donors in the direction of growth and a generally higher background of the donor doping in all the structure. Another difference between the SQW and MQW A samples is   in the energy of PL at $B$= 0 which is smaller in the case of MQW A by about 6~meV. Nevertheless, we apply the same solutions for the LLs as in the case of the SQW sample, adjusting appropriately the value of $E_0$. The oscillatory character of the transitions involving the lowest LL was discussed in detail in~\cite{JKunc_2010} and is of no concern here.

The inspection of Fig.~\ref{MQWData} shows that there are many transitions which are very well described by the theory. In particular, this refers to transitions between  LLs with higher indexes. On the other hand, there are many lines which are not theoretically reproduced. This could be related to at least two factors. First, the  QWs in a given sample need not to be identical and variations in the confining potential between different QWs could lead to a broadening of PL lines and even spectral separation of lines generated in different QWs. Second, a generally higher level of doping leads to a stronger disorder and makes   localization of photoexcited holes more probable than in the SQW sample. Then, some of transitions in the PL are related to free-to-bound or bound-to-bound transitions which are not described by the theory presented in this paper.


\subsection{Time-resolved photoluminescence}{\label{ TRPL}}
\begin{figure}[!htbp]
\centering
 \includegraphics[scale=0.35] {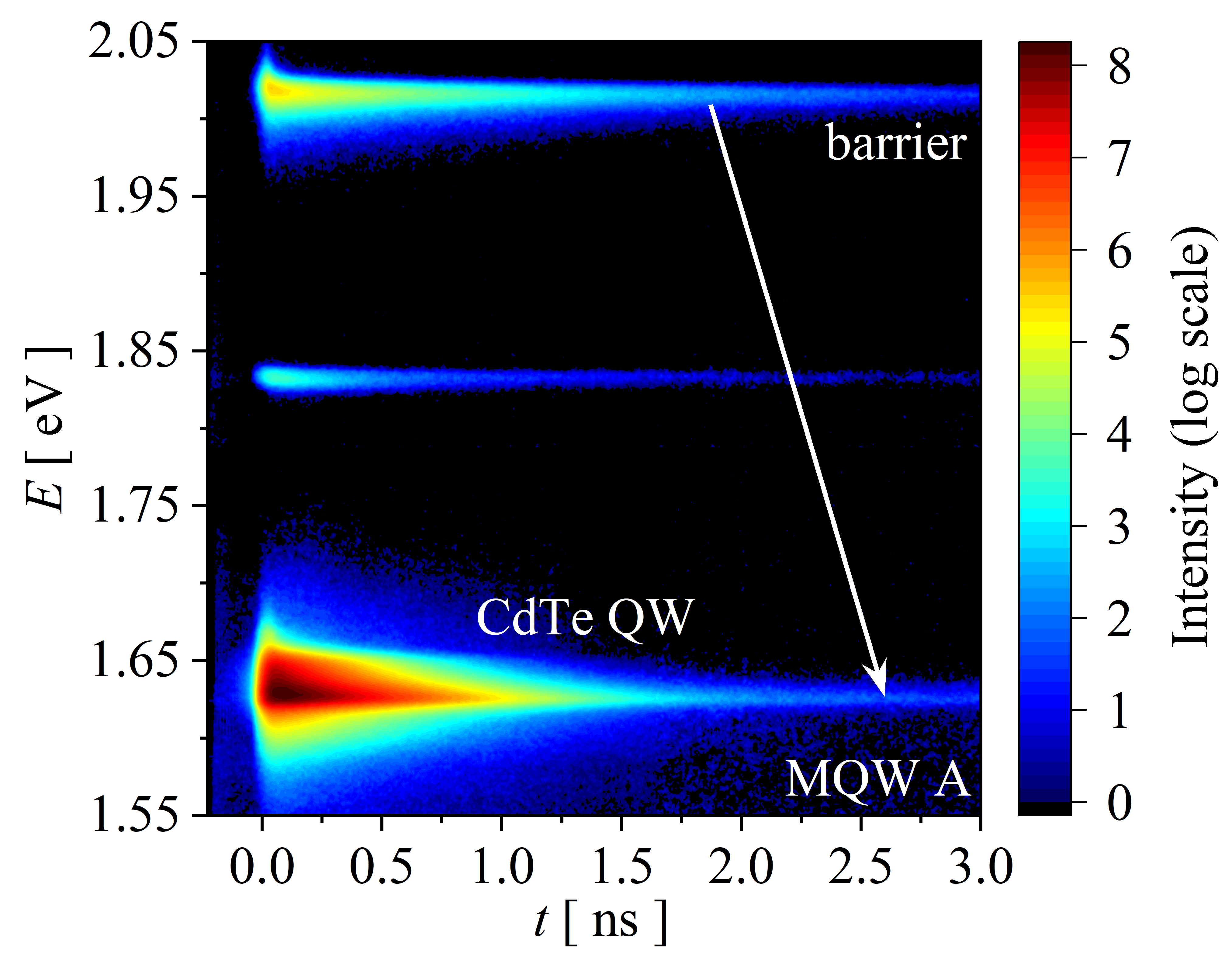}
\caption{\label{TRPL}A  false color map of the time-resolved PL for the \cdteQW\, MQW A sample shows the decay  of peak intensities, including peaks from the barrier, the supperlattice in the buffer and the QW. The white arrow symbolizes the proposed mechanism of  transfer of holes from the barrier to the QWs.  }
\end{figure}
Time-resolved measurements were carried out  in order to have a closer look on relaxation processes of photoexcited holes. The interest in this problem came from the fact that the relaxation of photoexcited carriers is typically very fast and it was interesting to understand, why we observed the PL from VB LLs with high indexes $n_v$ up to 5. According to Fig.~\ref{ValenceBand}, these holes are at several meV deep in the VB and at the first glance their presence there is not evident. 

According to the Boltzmann distribution at liquid helium temperature in a sample with holes as minority carriers, only the lowest energy hole levels should be occupied. Even if the generation of a hole at a high-energy level were to occur, such a hole should immediately relax, moving to one of the lowest levels. This would indicate that only transitions to these hole levels (with a low index $n_v$) would be visible on PL spectra. Due to the high concentration of electrons in the samples, we assume that electrons in the CB can occupy high-energy Landau levels in equilibrium and be involved in transitions. 

The  map of the time-resolved PL  for the MQW A sample presented in Fig.~\ref{TRPL} distinguishes three peaks: a peak from the CdTe QWs PL (i.e., the one whose magnetic field splitting is analyzed in this article), a peak from the superlattice PL  in the buffer layer and a peak from the barrier PL. The PL from the superlattice is much weaker than the other two and will be neglected in the following analysis. 

\begin{table}[hbt!]
\centering
\caption{Decay times (in ns) given by a  two-exponential fit to data presented in Fig.~\ref{TRPL}. Accuracy of values given in the table is estimated to be about 0.04~ns. }
\label{TimeResolved}
\begin{tabular}{c|cc|cc|}
\hline
 \multicolumn{1}{|c|}{Sample}                            & \multicolumn{2}{c|}{QW}                                    & \multicolumn{2}{c|}{Barrier}                               \\ \cline{2-5}
 \multicolumn{1}{|c|}{ } 			                & \multicolumn{1}{c|}{$\tau_1$  }         & $\tau_2$          & \multicolumn{1}{c|}{$\tau_1$ }              & $\tau_2$             \\ \hline
\multicolumn{1}{|c|}{SQW}    & \multicolumn{1}{c|}{$0.31 $} & $2.77 $ & \multicolumn{1}{c|}{$0.29 $} & $1.90 $ \\ \hline
\multicolumn{1}{|c|}{MQW A}  & \multicolumn{1}{c|}{$0.24 $} & $3.06 $ & \multicolumn{1}{c|}{$0.25 $} & $1.26 $ \\ \hline
\multicolumn{1}{|c|}{MQW B}  & \multicolumn{1}{c|}{$0.27 $} & $2.53 $ & \multicolumn{1}{c|}{$0.24 $} & $1.08 $ \\ \hline
\end{tabular}%
\end{table}

The decay of PL from the barrier and the QWs   has a two-stage character and the corresponding time constants are given in Table~\ref{TimeResolved}. They were determined by fitting of a two-exponential dependence to the data and gave a short $\tau_1$ and a long $\tau_2$ characteristic times. The short time is interpreted as a band-to-band recombination and it lasts a fraction of ns in each case.  This   swift decay   continues until all free carriers recombine. However, electrons and holes localize on defects or fluctuations of the electrostatic potential  and their recombination is a much longer process which corresponds to the second, longer decay, extending to a few ns.  

  We propose that during this time,  holes from the barrier tunnel into the QWs and a transfer of excitation from the barrier to the QWs occurs which is symbolized by the white arrow in Fig.~\ref{TRPL}.   The decay of the PL from the QWs shows also a two-stage character, which is consistent with the proposed mechanism  of a transfer of holes from the barrier: after the rapid recombination of free holes photoexcited directly in the QWs, barrier holes tunnel into the QWs  and the PL can still occur, especially to the "deeper" hole levels which are energetically aligned with holes' levels in the barrier. 

We note that   the fast relaxation time $\tau_1$ is in all cases almost identical which could be expected for electron - hole recombination  in very similar materials (CdTe QW and \cmt barriers). On the other hand, a difference of $\tau_2$ between the QW and the barrier could be interpreted as a time needed for tunneling.

\section{Conclusions}{\label{Conclusions}}

This paper concerns a near band-gap magnetophotoluminescence from  single and multiple CdTe/(CdMg) QWs modulation-doped with iodine. Our work consists of two parts: experimental and theoretical. The experimental part describes polarization-resolved PL measurements at liquid helium temperatures  and magnetic fields up to 9~T and zero-field time-resolved PL measurements. The theoretical part comprises  calculations of the energy of LLs in the CB and VB. The energy of CB LLs  are determined within the 3LM assuming a coupling of the $\Gamma_6^c$ levels in the CB with the $\Gamma_8^v$ and $\Gamma_7^v$ levels in the VB. The VB is described by the Luttinger Hamiltonian for the $\Gamma_8^v$ band adapted for the case of a QW. 

Preliminary, the energy of transitions is calculated with taking into account the selection rules derived on the assumption that the VB LLs are mixtures of the $\Gamma_8^v$ Bloch functions   and   LL wave functions in the CB are proportional to the $\Gamma_6^c$ $S$- symmetry Bloch function. The selection rules derived within these assumptions (presented in Eq.~\ref{SelectionRules}) lead to quite a good agreement between the theoretical description and the  experimental data with exception of strong transitions which apparently ``escaped'' from this theoretical model. To deal with this problem, we propose that these  transitions can be  found in the theoretical description if one notes that next to an admixture of $\Gamma_8^v$ and $\Gamma_7^v$  wave functions to the CB there is an accompanying admixture of $\Gamma_6^c$ from the CB to the $\Gamma_8^v$ in the VB.  

We also carried out time-resolved PL measurements at $B=0$ which helped us to understand, why we observed transitions involving LLs in the VB with high numbers. We propose that   photoexcited barriers is a reservoir of long-lived holes tunneling to high-index LLs in the QW VB. 

The novelty of the present paper lies in application of the \kp theory to describe the PL from \cdteQW\, QWs which seems not to be presented before. We also propose the enlargement of the set of selection rules  of free-to-free recombination derived in Ref.~\cite{MKubisa_2003} resulting from mixing of the conduction and valence bands.

\appendix
\section{\label{KacmanZawadzki}}
There are two wave functions related to the $n^{\mbox{\small{th}}}_c$ LL in the CB which originate from $\phi_{n_c} u_1$ and $\phi_{n_c} u_2$ due to mixing of the CB with the VB. These function were named $\Psi_{{n_c}{k_y}{k_z}-}(\boldsymbol{r})$ and $\Psi_{{n_c}{k_y}{k_z}+}(\boldsymbol{r})$ and for a  bulk zinc-blende material at the $\Gamma$ point of the Brillouin zone  are   the following (we keep here the original notation of Ref.~\cite{PKacman_1971}; $\Phi_n$ are harmonic oscillator functions denoted by $\phi_{n_c}$ in the present paper):

\begin{widetext}

\begin{align}
\begin{aligned}
&\Psi_{n k_y k_z -}(r) = \mathrm{exp}(i k_y y + i k_z z) \times \Bigl\{\Bigl[i a_- \Phi_n S + \frac{b_- + c_- \sqrt{2}}{2} \times \\ 
&\times \Bigl(\frac{\hbar \omega_c (n + 1)}{D_{n-}}\Bigr)^{1/2} \Phi_{n + 1} R_- - \frac{b_- - c_- \sqrt{2}}{2} \Bigl(\frac{\hbar \omega_c n}{D_{n-}}\Bigr)^{1/2}  \Phi_{n - 1} R_+ + \\ 
&+ c_-\Bigl(\frac{\hbar^2 k_z^2 / (2m_0^*)}{D_{n-}}\Bigr)^{1/2} \Phi_n Z \Bigr] \downarrow + \\ 
&+ b_-\Bigl[\Bigl(\frac{\hbar^2 k_z^2 / (2m_0^*)}{D_{n-}}\Bigr)^{1/2}\Phi_n R_- - \Bigl(\frac{\hbar \omega_c n}{D_{n-}}\Bigr)^{1/2} \Phi_{n-1} \frac{Z}{\sqrt{2}} \Bigr] \uparrow \Bigr\}
\end{aligned}
\end{align}

\begin{align}
\begin{aligned}
&\Psi_{n k_y k_z +}(r) = \mathrm{exp}(i k_y y + i k_z z) \times \Bigl\{\Bigl[i a_+ \Phi_n S + \frac{b_+ - c_+ \sqrt{2}}{2} \times \\ 
&\times \Bigl(\frac{\hbar \omega_c (n + 1)}{D_{n+}}\Bigr)^{1/2} \Phi_{n + 1} R_- + \frac{b_+ - c_+ \sqrt{2}}{2} \Bigl(\frac{\hbar \omega_c n}{D_{n+}}\Bigr)^{1/2}  \Phi_{n - 1} R_+ + \\ 
&+ c_+\Bigl(\frac{\hbar^2 k_z^2 / (2m_0^*)}{D_{n+}}\Bigr)^{1/2} \Phi_n Z \Bigr] \uparrow + \\ 
&+ b_+\Bigl[\Bigl(\frac{\hbar^2 k_z^2 / (2m_0^*)}{D_{n+}}\Bigr)^{1/2}\Phi_n R_+ - \Bigl(\frac{\hbar \omega_c (n+1)}{D_{n+}}\Bigr)^{1/2} \Phi_{n+1} \frac{Z}{\sqrt{2}} \Bigr] \downarrow \Bigr\}
\end{aligned}
\end{align}
\end{widetext}
where for energies $\lambda_{\pm} \ll (2/3) \Delta$ the coefficients are

\begin{align}
\begin{aligned}
&a_{\pm}^2 = \frac{\lambda_{\pm}}{\epsilon_0 + 2 \lambda_{\pm}},  
&b_{\pm}^2 = \frac{1}{3}\frac{\epsilon_0 + \lambda_{\pm}}{\epsilon_0 + 2 \lambda_{\pm}},  \\
&c_{\pm}^2 = \frac{2}{3}\frac{\epsilon_0 + \lambda_{\pm}}{\epsilon_0 + 2 \lambda_{\pm}},  \:\:\:\: 
&R_{\pm} = \frac{1}{\sqrt{2}}(X \pm i Y).
\end{aligned}
\end{align}

$\epsilon_0$ is the band gap, $D_{n\pm}$ and $\lambda_{\pm}$ are defined as follows:

\begin{align}
\begin{aligned}
\lambda(n, k_z, \pm) &= -\frac{\epsilon_0}{2} + \Bigl[\Bigl(\frac{\epsilon_0}{2}\Bigr)^2 + \epsilon_0 D(n, k_z, \pm)\Bigr]^{1/2}, \\
D(n, k_z, \pm) &= \hbar \omega_c \Bigl(n+\frac{1}{2}\Bigr) + \frac{\hbar^2 k_z^2}{2m_0^*} \mp \frac{1}{2} \mu_B |g_0| H.
\end{aligned}
\end{align}

To shorten the description of these functions used in the discussion of our experimental results, we present the above equations in a more concise form:
\begin{align}
\begin{aligned}
 &\Psi_{{n_c}{k_y}{k_z}-}(\boldsymbol{r}) = \exp(ik_y y+ik_z z)\times \\ 
 &\left\{\left[ ia_{-}\phi_{n_c} S + \alpha_1 \phi_{n_c+1}(X-iY)\right. \right. \\
 &\left. -  \alpha_2 \phi_{n_c-1}(X+iY) + \alpha_3 \phi_{n_c}Z \right] \downarrow \\
 &+\left. \left[ \alpha_4 \phi_{n_c}(X-iY) - \alpha_5 \phi_{n_c-1}Z\right]\uparrow\right\}
  \label{Psi-}
\end{aligned}
\end{align}

\begin{align}|
\begin{aligned}
 &\Psi_{{n_c}{k_y}{k_z}+}(\boldsymbol{r}) = \exp(ik_y y+ik_z z)\times \\
& \left\{\left[ ia_{+}\phi_{n_c} S - \beta_1 \phi_{n_c+1}(X-iY)\right. \right.\\
& \left. +  \beta_2 \phi_{n_c-1}(X+iY) + \beta_3 \phi_{n_c}Z \right] \uparrow \\
 &-\left. \left[ \beta_4 \phi_{n_c}(X+iY) - \beta_5 \phi_{n_c+1}Z\right]\downarrow\right\}
 \label{Psi+}
\end{aligned}
\end{align}

The definitions of coefficients in the above equations can be found in~\cite{PKacman_1971}. From the point of view of this work, it is important to note that the coefficients $\alpha_1, \alpha_2, \beta_1$ and $\beta_2$ are proportional to $\sqrt{\hbar \omega_c}$. This indicates that admixture of the wave functions from the VB, which is accompanied by admixture to the $n^{\mbox{\small{th}}}_c$ LL of neighboring LLs (with indexes $n_c+1$ and $n_c -1$) increases with the magnetic field.

\vspace{-0.5cm}
\begin{acknowledgements}

Fruitflul discussions with   Jan Suffczyński are kindly acknowledged.

This research was partially supported by the Polish National Centre grant UMO-2019/33/B/ST7/02858, by the “MagTop” project (FENG.02.01-IP.05-0028/23) carried out within the “International Research Agendas” programme of the Foundation for Polish Science co-financed by the European Union under the European Funds for Smart Economy 2021-2027 (FENG). Publication subsidized from the state budget within the framework of the programme of the Minister of Science (Polska) called Polish Metrology II project no. PM-II/SP/0012/2024/02, amount of grant 944,900.00 PLN, total value of the project 944,900.00 PLN.

The work was supported by the European Union through ERC-ADVANCED grant TERAPLASM (No. 101053716). Views and opinions expressed are, however, those of the author(s) only and do not necessarily reflect those of the European Union or the European Research Council Executive Agency. Neither the European Union nor the granting authority can be held responsible for them. We also acknowledge the support of "Center for Terahertz Research and Applications (CENTERA2)" project (FENG.02.01-IP.05-T004/23) carried out within the "International Research Agendas" program of the Foundation for Polish Science co-financed by the European Union under European Funds for a Smart Economy Programme.

\end{acknowledgements}
 \bibstyle{plain}
\bibliography{bibliographyJL}

\end{document}